\documentclass[12pt, twoside]{article}   	
\usepackage{geometry}                		
\usepackage{graphicx}	 
\usepackage{amsmath,amsfonts,amsbsy} 
\usepackage{slashed} 
\usepackage{authblk}

\usepackage[scr=boondoxo,scrscaled=1.05]{mathalfa}

\newcommand{\ee}{\mathrm{e}} 
\newcommand{\ve}{\varepsilon}
\newcommand{\euv}{\mathrel{\mathop{=}\limits_{^{^\mathrm{UV}}}}}
\newcommand{\pk}{p{\cdot}k}

\newcommand{\xk}{x{\cdot}k}
\newcommand{\laser}{\mathcal{A}}
\newcommand{\lasers}{\laser^{*}}

\newcommand{\Prop}[2]{\mathrm{P}{#1}_{#2}}
\newcommand{\Props}[1]{\mathrm{P}({#1})}
\newcommand{\Ab}{\mathrm{A}}
\newcommand{\Em}{\mathrm{E}}
\newcommand{\In}{\mathrm{I}}
\newcommand{\Out}{\mathrm{O}}

\newcommand{\sigmauv}{\Sigma^{^{\mathrm{UV}}}}
\newcommand{\psl}{\slashed{p}}
\newcommand{\ksl}{\slashed{k}}
\newcommand{\laserforslash}{\mathcal{A}\,\,}
\newcommand{\lasersl}{\slashed{\laserforslash}\!\!} 
\newcommand{\ssl}{\slashed{s}}
\newcommand{\laserssl}{\slashed{\laser^*}}
\newcommand{\mstar}{\mathscr{M}}
\newcommand{\smstar}{\mathfrak{m}}
\newcommand{\mstarsl}{\slashed{\mstar}}
\newcommand{\sigmasl}{{\Sigma}_{_{\!\!\mstarsl}}}
\newcommand{\sigmasluv}{\sigmasl^{^{\mathrm{UV}}}}

\newcommand{\sigmase}{\Sigma_{_{\mathrm{s\_\nobreak\hspace{.05em}se}}}}
\newcommand{\sigmaseuv}{\Sigma^{^{\mathrm{UV}}}_{_{\mathrm{s\_\nobreak\hspace{.05em}se}}}}
\newcommand{\sigmafse}{\Sigma_{_{\mathrm{f\_\nobreak\hspace{.05em}se}}}}
\newcommand{\sigmafseuv}{\Sigma^{^{\mathrm{UV}}}_{_{\mathrm{f\_\nobreak\hspace{.05em}se}}}}
\newcommand{\sigmama}{\Sigma_{_{\smstar^2}}}
\newcommand{\sigmamauv}{\Sigma^{^{\mathrm{UV}}}_{_{\smstar^2}}}

\newcommand{\sigmain}{\Sigma_{_{\mathrm{in}}}}
\newcommand{\sigmainuv}{\Sigma^{^{\mathrm{UV}}}_{_{\mathrm{in}}}}

\newcommand{\sigmaoutuv}{\Sigma^{^{\mathrm{UV}}}_{_{\mathrm{out}}}}
\newcommand{\deltauv}{\Delta^{^{\!\!\mathrm{UV}}}}

\newcommand{\intfsD}{\int{\bar{}\kern-0.45em d}^{\,D\!}s\,}
\newcommand{\intfsDuv}{\int_{_\mathrm{UV}}{\bar{}\kern-0.45em d}^{\,D\!}s\,}

\newcommand{\intfp}{\int{\bar{}\kern-0.45em d}^{\,4}p\,}

\begin{document} 

 \title{The simplest description of  charge propagation in a strong background}
 \author{Martin Lavelle\\ \texttt{m.lavelle@plymouth.ac.uk} \and David McMullan\\ \texttt{d.mcmullan@plymouth.ac.uk}}
 \affil{Centre for Mathematical Sciences\\University of Plymouth\\Plymouth, PL4 8AA, UK}
 \date{7$^\mathrm{th}$ November, 2020 }

\maketitle

 \begin{abstract} 
Exploiting the gauge freedom associated with the Volkov description of a charge propagating in a plane wave background, we identify a new  type of gauge choice which significantly simplifies the theory. This allows us to   develop a compact description of the propagator for both scalar and fermionic matter, in a circularly polarised background. It is shown that many of the usually observed structures are gauge artefacts. We then analyse  the full ultraviolet behaviour of the one-loop corrections for such charges. This enables us to identify and contrast the different renormalisation prescriptions needed for both types of matter. 
\end{abstract}
 
\section{Introduction} 

Very early  in the development of quantum electrodynamics, QED, it was understood that the interaction of light with matter was best described in a way that introduced extra, unphysical, degrees of freedom,~\cite{Dirac1}\cite{Fermi1}. The expected two components of the photon, at each spacetime point, were embedded  in the four components  of the vector potential, as these were  needed to formulate the interaction with matter.  The recovery of physical results then followed from the gauge invariance of QED. Gauge fixing allows for a direct  recovery of the physical dynamics of the theory. This is the case both for interactions in the vacuum and  in a background. For example, in the Volkov description of the propagation of matter through a plane wave background, \cite{Volkov:1935zz}, there is also an implicit  gauge fixing for the background field.

Counting degrees of freedom in such gauge theories is complicated by the Lorentzian signature of spacetime. The naive expectation would be that two gauge fixing conditions are needed to remove the two extra degrees of freedom, but in practice a single covariant gauge suffices to define photon propagators and hence S-matrix elements. Unitarity arguments can then show that suitably defined cross-sections between appropriate states correspond to physical results with the correct degrees of freedom.

For propagation in a background, the gauge freedom in describing the background potential is, as noted above, often implicit in the formalism. That is, the explicit form of the  potential  implies that  a gauge fixing condition has been used.  So for the plane wave situation described by the Volkov solution, the scalar product of the null momentum, pointing along the beam, with the background potential vanishes. This is essentially a light cone gauge choice  for the potential. There is still some residual gauge freedom in the choice of the background potential, but there is no fundamental requirement  for adding an additional gauge fixing condition on the background.  

However, the Volkov solution is very complicated and disentangling physics from flotsam is a challenge. In this paper we will argue that a specific  choice of additional gauge fixing on the background can significantly simplify the description of both the classical and quantum propagation of a charge through the background. We shall see that this holds for both weak and strong backgrounds, and leads to clear renormalisation conditions on the fields and physical parameters. This will be shown for both scalar and fermionic matter, and in this way we will be able to highlight and contrast some of the simple  results found here for the renormalisation of both theories through the use of our additional gauge fixing condition on the background.

The approach taken here is perturbative in the strength of the background, and in that way we can build upon the familiar and precise language of perturbative quantum field theory. We will thus be able to explicitly introduce counterterms and renormalise using standard field theory constructions. That this can be done for both types of matter and for both weak and strong backgrounds, points to the great utility of imposing our additional gauge fixing condition on the background. This paper will take the background to be circularly polarised, as that choice will lead to the simplest possible expressions for the propagator, especially in the context of scalar matter.

The plan of this paper is to first discuss, in section~2, the background gauge freedom  and introduce the additional, momentum gauge choice on it. Then, in section~3,  we couple the background to scalar matter. The great utility of the momentum gauge is demonstrated here as  we will be able to present a full and self contained account of the propagation, and its one-loop ultraviolet  corrections, of such matter in both a weak and strong background. As well as being of great interest in its own right, this will allow us to introduce some of the key arguments and notation that will then be refined when we treat the case of fermionic matter in the rest of the paper.

In section~4 we consider fermionic matter, and introduce the key ingredients needed to describe its interaction with the background. Then, in section~5, we derive the full tree level fermionic propagator in the background, and see again how the additional momentum gauge choice greatly simplifies this derivation. The one-loop corrections to the propagator in a weak background are presented in section~6, and these are extended to the full, strong background, one-loop calculations in section~7. Armed with these results, we then go on to discuss the renormalisation of both the scalar and fermionic theories in section 8. We then conclude this paper in section 9, where we also discuss how the approach taken here can be extended to other polarisation choices for the background.  Some key technical results are given in appendices.

\section{Background field gauge freedom} 
The real classical  potential, $A_{\mathrm{c}}^\mu$  ,   describing a circularly polarised  background is most conveniently written as the sum of two conjugate fields: 
\begin{equation}\label{eq:potential}
  A_{\mathrm{c}}^\mu=\laser^\mu+{\lasers}^\mu\,,
\end{equation}
where the complex potential is given by
\begin{equation}\label{eq:A_circular}
  \laser^\mu=\tfrac12e\big(a_1^\mu+ia_2^\mu\big)\ee^{-i\xk}\,.
\end{equation}
Here $k^\mu$ is the null momentum characterising the plane wave background, $a_1^\mu$ and $a_2^\mu$ are  orthogonal, real, spacelike  vectors which satisfy the common normalisation condition that  $a^2:=a_1\cdot a_1=a_2\cdot a_2<0$.  This complex potential also  satisfies the null gauge condition
\begin{equation}\label{eq:null_gauge}
  k\cdot\laser=0\,,
\end{equation}
which is equivalent to the two real conditions that $k\cdot a_1=k\cdot a_2=0$. 

It should be noted that choosing  a particular direction along which the background points introduces both a directional, $x$, and momentum, $k$, dependence  to the potential $\laser^\mu$, as is clear from the final term in (\ref{eq:A_circular}). We shall soon see, though, that this potential is essentially the background interaction term in a perturbative approach to the system, and it is for that reason that we suppress its explicit dependence on these variables.  However, as discussed in \cite{Lavelle:2019lys}, and shown later here at all orders in perturbation theory, this  will still lead to a multiplicative, momentum space renormalisation procedure.

It is important to also note that there is still a residual gauge freedom in the potential, since  $\laser^\mu+\Lambda k^\mu$ also satisfies the  null gauge condition (\ref{eq:null_gauge}), for arbitrary $\Lambda$ with the same spatial dependence as $\laser^\mu$. In terms of the real potentials, this gauge freedom is $a_1^\mu\to a_1^\mu+\Lambda_1 k^\mu$ and $a_2^\mu\to a_2^\mu+\Lambda_2 k^\mu$, where then $\Lambda=\frac12e(\Lambda_1+i\Lambda_2)\ee^{-i\xk}$.

It is helpful to be a bit more explicit about this residual gauge freedom. If we write $k^\mu=(k^0,0,0,k^0)$, then all the conditions on the real potentials are satisfied by writing
 \begin{equation}\label{eq:a1a2}
    a_1=\begin{pmatrix} 
      0\\\alpha\\\beta\\0
    \end{pmatrix}+\Lambda_1 \begin{pmatrix}
      k^0\\0\\0\\k^0
    \end{pmatrix}\qquad\mathrm{and}\qquad
 a_2=\pm\begin{pmatrix}
      0\\\beta\\-\alpha\\0
    \end{pmatrix}+\Lambda_2 \begin{pmatrix}
      k^0\\0\\0\\k^0
    \end{pmatrix}\,,
  \end{equation} 
where the common amplitude normalisation is $a^2=-\alpha^2-\beta^2$, and the sign ambiguity reflects left or right polarisation choices, as discussed in~\cite{Lavelle:2017dzx}. 

It is tempting to think of the two parameters $\alpha$ and $\beta$ in (\ref{eq:a1a2}) as the natural representation of the true degrees of freedom for the background. Indeed, for light by light scattering,  such an identification is sensible.  But it is not necessarily the best representation of the background when matter is present. We now introduce a new characterisation of the true degrees of freedom  that is much better suited to calculations involving a charge propagating through the background.

Consider a charge of mass $m$ that  has associated with it  a timelike momentum $p^\mu$ describing its propagation through the background. This momentum may, or may not, be taken to be on-shell. But, given that $k^\mu$ is the fixed null momentum associated with the background, we can ensure that $p\cdot k\ne0$. What is more, in this plane wave description,  the momentum $p$ is interpreted as an external momentum and thus is not integrated over in any loop calculation associated with the propagation of the charge. So we are able to ensure that $p\cdot k$ will never vanish in both the tree level propagator and its loop corrections.  

We now impose an additional  gauge condition on the background potential by requiring that, as well as (\ref{eq:null_gauge}), we have 
\begin{equation}\label{eq:mmt_gauge}
  p\cdot\laser=0\,.
\end{equation}
In terms of the explicit representation (\ref{eq:a1a2}), this momentum gauge condition fixes the residual gauge freedom so that
\begin{equation}
  \Lambda_1=\frac{p_1\alpha+p_2\beta}{p\cdot k}\qquad\mathrm{and}\qquad\Lambda_2=\pm\frac{p_1\beta-p_2\alpha}{p\cdot k}\,.
\end{equation}
For example, if the charge was static, or moving solely along the $z$-axis, so that  $p=(p_0,0,0,\lambda p_0)$ with $|\lambda|<1$, 
then $\Lambda_1=\Lambda_2=0$, and we have the very natural representation mentioned earlier with $a_1^\mu$ and $a_2^\mu$ only having  components in the transverse directions to the background. 

But now, suppose the particle was moving along the $x$-axis. The momentum can then be written as $p=(p_0,p_1,0,0)$, and we impose the timelike requirement  that  $p^2=p^2_0-p^2_1>0$. Then we find, from (\ref{eq:a1a2}), that
\begin{equation}\label{eq:a1a2p1}
    a_1=\begin{pmatrix} 
      \dfrac{p_1\alpha}{p_0}\\[0.2cm]\alpha\\\beta\\[0.2cm]\dfrac{p_1\alpha}{p_0}
    \end{pmatrix}\qquad\mathrm{and}\qquad
 a_2=\pm\begin{pmatrix}
      \dfrac{p_1\beta}{p_0}\\[0.2cm]\beta\\-\alpha\\[0.2cm]\dfrac{p_1\beta}{p_0}
    \end{pmatrix}\,.
  \end{equation}
We shall see that these simple examples, and their  full timelike extensions,  give a  computationally efficient   way to characterise the  background field for such a propagating charge. 

\bigskip

Note that the above mentioned  static class of representations of the background potential could also be characterised by the additional light cone condition that $\bar{k}\cdot \laser=0$, where $\bar{k}$ points along  the dual light cone direction: $\bar{k}^\mu=(k^0,0,0,-k^0)$. This  choice also ensures that $k\cdot\bar{k}\ne0$ as it is just the $\lambda\to-1$ limit of our earlier static class. In applications to light by light scattering, there is also great utility in this additional gauge choice, see for example \cite{Dinu:2013gaa}, \cite{Meuren:2013oya} and \cite{King:2015tba}. However, in the context of particle propagation, we shall see that focusing on the light cone structure obstructs the rich interplay between the lightlike  background and the timelike particle dynamics inherent in this system. Exploiting this will lead to significant computational advantages and clearer physical insight in to this complex but important system.  

\medskip

Before concluding this general introduction to the kinematics of our system, it is worth noting that the above discussion of the momentum gauge choice assumes that it is sensible to talk of the charge as having a given momentum, $p$. Obviously, in the context of scattering, the momentum will change. Any such measurable scattering is not an ambiguity in the formalism, and the momentum gauge can still be used for at least, say, the incoming particle.  However, even in the context of simple charge propagation,  with no additional external interactions, the background itself obscures any idea of an unambiguous particle momentum.

So, although we have characterised the charge as having momentum $p$, the fact that it is propagating in a background means that the actual momentum is ambiguous. More precisely, we should allow for its momentum to be of the form $p+nk$, where the integer $n$ counts the number of absorptions from the background minus the number of emissions degenerate to the background.

It is important, though,  to note that this change in the momentum will not affect  the overall gauge fixing conditions being proposed here since, from (\ref{eq:null_gauge}) and (\ref{eq:mmt_gauge}), we also have $(p+nk)\cdot\laser=0$, for all possible values of $n$. 

\medskip
We now begin our analysis of the propagation of matter through this background. Although our primary interest is in fermionic matter, we shall start with the much simpler case of scalar matter. Our choice of polarisation and momentum gauge now becomes particularly effective, and the transition to intense backgrounds will be almost immediate. This will be a good test case and help motivate the key definitions needed for the more complex fermionic structures that will be the main focus of this paper.

\section{Scalar matter} \label{sec:scalar_matter}
The quadratic nature  of the Lagrangian for scalar QED  means that the matter interacts with  photons via either a three or four point vertex. The Feynman rules for these vertex contributions are given by the truncated\footnote{See the discussion on page 90 of \cite{Sterman:1994ce} concerning such truncated lines and possible representations.} diagrams, i.e., Green functions with external lines removed as signified by the small bars on them:
\begin{equation}\label{eq:s_f_rules}
  \raisebox{-0.6cm}{\includegraphics{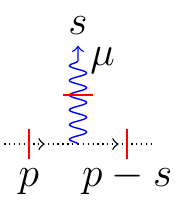}}=-ie(2p_\mu-s_\mu)\quad\mathrm{and}\quad\raisebox{-0.6cm}{\includegraphics{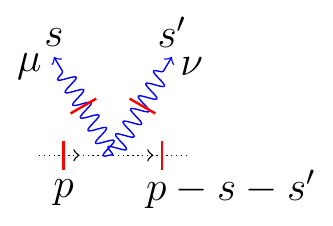}}=2ie^2g_{\mu\nu}\,.
\end{equation}
Dotted lines are used here to represent the scalar propagators while wavy lines correspond to the photons. 
These rules are equivalent to those derived in, for example, section 6-1-4 of \cite{Itzykson:1980rh}. 

When the photon is taken to be degenerate with the plane wave background, these vertex terms can be contracted with suitably normalised products of the background potential, (\ref{eq:potential}), to give the background  interactions with the scalar matter in terms of either absorptions 
\begin{equation}\label{eq:s_ab}
  \raisebox{-0.6cm}{\includegraphics{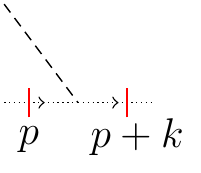}}:=-i2p\cdot\laser\,,
\end{equation}
or emissions
\begin{equation}\label{eq:s_em}
  \raisebox{-0.6cm}{\includegraphics{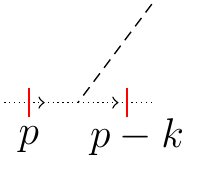}}:=-i2p\cdot\lasers\,,
\end{equation}
along with the various mixed absorption or emission, seagull interactions
\begin{equation}\label{eq:s_gull}
  \raisebox{-0.6cm}{\includegraphics{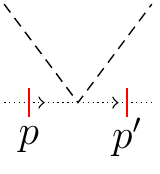}}:=\tfrac12 A_{\mathrm{c}}^\mu 2ie^2g_{\mu\nu}A_{\mathrm{c}}^\nu\,.
\end{equation}

Our choice of momentum gauge, (\ref{eq:mmt_gauge}), then greatly simplifies these interactions as both the absorption, (\ref{eq:s_ab}), and emission, (\ref{eq:s_em}), interactions vanish. In addition, our choice of circular polarisation, (\ref{eq:A_circular}), means that both  $\laser\cdot\laser=0$ and $\lasers\cdot\lasers=0$, and we quickly see that the only surviving seagull interaction is the momentum conserving, $p'=p$, one with Feynman rule
\begin{equation}
  \tfrac12 A_{\mathrm{c}}^\mu 2ie^2g_{\mu\nu}A_{\mathrm{c}}^\nu=2i\lasers{\cdot}\laser=ie^2a^2\,.
\end{equation}
From the discussion preceding (\ref{eq:null_gauge}), we see that the background amplitude satisfies a spacelike normalisation condition, $a^2<0$. It is thus useful to introduce  the positive scalar quantity, $\smstar>0$, defined by
\begin{equation}\label{eq:mstar2}
  \smstar^2=-e^2 a^2\,.
\end{equation}
Hence we see that, by using the momentum gauge (\ref{eq:mmt_gauge}), the sole  surviving interaction of the scalar matter with the circularly polarised background is given by the simple seagull term: 
\begin{equation}\label{eq:v_int}
  \raisebox{-0.6cm}{\includegraphics{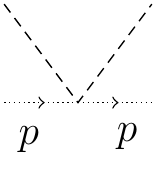}}=\Props{m^2}\big(-i \smstar^2\big)\Props{m^2}\,.
\end{equation}
Note that  we have introduced in this last expression a useful, compact  notation for the scalar propagator that emphasises its mass dependence, so that
\begin{equation}\label{eq:props}
  \Props{m^2}:=\frac{i}{p^2-m^2+i\epsilon}\,.
\end{equation} 
Two such seagull interactions are then given by 
\begin{equation}\label{eq:2seagulls}
  \raisebox{-0.6cm}{\includegraphics{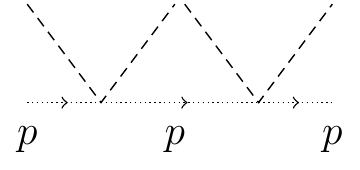}}=\Props{m^2}\big(-i \smstar^2\big)\Props{m^2}\big(-i \smstar^2\big)\Props{m^2}\,.
\end{equation}

From this we see that multiple interactions with the background are now simple products of these seagulls.  So $r\ge0$ such interactions  can be represented as
\begin{equation}\label{eq:r_v_int}
  \raisebox{-0.6cm}{\includegraphics{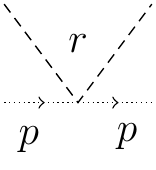}}:=\Props{m^2}\big(-i \smstar^2\Props{m^2}\big)^r\,.
\end{equation}
Note that $r=0$ here corresponds to the scalar propagator $\Props{m^2}$. When $r=1$ we will often omit the label, as in (\ref{eq:v_int}). 

Summing this last result over all possible values for $r$ will then describe the physical propagation of the scalar charge through the background. 
It is important to be able to distinguish the resulting all orders propagation  from the usual, perturbative, vacuum propagation of the charge. The convention adopted for a long time, as can be seen in section~105 of \cite{Berestetsky:1982aq}, was to use a thicker line to represent the propagator in the background. However, there has been a trend in recent years to make visually clearer the distinct types of propagation being considered in these complex systems. This  can be seen in, for example, Figure~6 in \cite{Bamber:1999zt} and Figure~1 in \cite{Heinzl:2009nd}, where a double line was used to distinguish propagation in the background.  

We hence  define the scalar double line propagator of momentum $p$, in the momentum gauge (\ref{eq:mmt_gauge}), by
\begin{equation}\label{eq:scalar_doubleline}
  \raisebox{-0.05cm}{\includegraphics{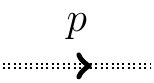}}=\sum_{r=0}^{\infty}
  \raisebox{-0.6cm}{\includegraphics{scalar_rr.pdf}}\,.
\end{equation} 
Thus we have the simple result, which follows immediately from (\ref{eq:r_v_int}), that
\begin{equation}\label{eq:scalar_mass_shift}
  \raisebox{-0.05cm}{\includegraphics{scalar_doubleline.pdf}}=\Props{m^2+\smstar^2}\,.
\end{equation}
So, from this strong field summation of the interactions with the background, it is clear that the only impact of the background on the scalar particle is that its mass has increased to $m_*^2:=m^2+\smstar^2$. This is surprisingly simple. Normally this propagator would involve an infinite sum over different poles (sidebands) and, most strikingly, break translational invariance. None of these complications are present due to our momentum gauge choice.

\bigskip
 
Having constructed the tree level propagator for the scalar particle in our  momentum gauge fixed  background, we now consider its one-loop corrections and the associated ultraviolet structures. This will require an additional gauge fixing choice to be made, but now its role is to allow for the construction of the  photon's propagator within the loop, rather than the description of the  charge's propagation through the background. 

In order to understand the impact of this gauge choice on the one-loop structure, we will take the  photon's propagator, $D_{\mu\nu}(s)$, to be in the full Lorentz class of gauges, as  given by
\begin{equation}
    D_{\mu\nu}(s)=\frac{-i}{s^2+i\epsilon}\Big(g_{\mu\nu}+(\xi-1)\frac{s_\mu s_\nu}{s^2}\Big)\,.
  \end{equation}
We recall that this class includes the Feynman gauge, where $\xi=1$, and Landau (or Lorenz)  gauge when $\xi=0$.  

In addition to gauge fixing, loops require a method to regularise the ultraviolet sector, and for that we adopt  dimensional regularisation. So we take the  spacetime dimension to be $D=4-2\varepsilon$, with $\varepsilon>0$, and introduce a mass scale, $\mu$, to maintain the canonical dimensions for the loop integral and renormalised fields.

As  expected, loop corrections to the scalar propagator in the background  will contain ultraviolet divergences, but these are now worse than those encountered in the fermionic theory. In addition to the logarithmic divergences, we now also get quadratic  ones. The great attraction to using dimensional regularisation is that it deals with these polynomial types of ultraviolet structures in a very efficient way by putting them equal to zero. But some care is needed in doing that as we  can also encounter other classes of divergences that can interfere with this prescription.

The simplicity of dimensional regularisation  can be seen most dramatically in the seagull loop diagram given by
\begin{equation}\label{eq:tadpole}
  \raisebox{-1.12cm}{\includegraphics{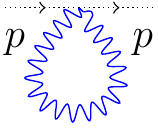}}=\Props{m^2}\,2(D+\xi-1)e^2\mu^{2\ve}\!\!\intfsD\frac1{s^2}\,\Props{m^2}\,.
\end{equation}
By simple power counting, we see that the loop integral in (\ref{eq:tadpole}) diverges quadratic in the large $s$, ultraviolet sector. But, within dimensional regularisation, this integral can be evaluated for some $D<4$ and then analytically continued back to four dimensions. The end result is that the integral vanishes. Thus there is no contribution to the propagator from the one-loop term (\ref{eq:tadpole}).

\medskip

The non-vanishing one loop correction to the scalar propagator is thus given by the self-energy term
\begin{equation}\label{eq:scalar_se}
  \raisebox{-0.6cm}{\includegraphics{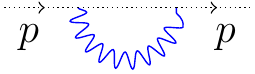}}=\Props{m^2}\big(-i\sigmase\big)\Props{m^2}\,,
\end{equation} 
where the scalar's self-energy is given at one-loop by
\begin{align}\label{eq:scalar_sigma_se}
\begin{split}
  -i\sigmase=-e^2\mu^{2\ve}\!\!\!\intfsD
  &\bigg(\frac{(p+s)^2}{(s^2-m^2)(s-p)^2}\\&+(\xi-1)\frac{(p^2-s^2)(p^2-s^2)}{(s^2-m^2)(s-p)^2(s-p)^2}\bigg)\,,
\end{split}
\end{align} 
and we have suppressed for brevity the $i\epsilon$ prescription for the poles of the various propagators within the loop.

Again we see  quadratic divergences here coming from the $s^2$ term in the first numerator and the $s^2s^2$ factor in the second. But in addition, the second term  also produces a subleading,  logarithmic  divergence. Factorising the numerators so as to cancel terms in the denominators leads to finite parts, which we are not considering in this paper, plus the ultraviolet divergent terms:
\begin{equation}
  -i\sigmase\euv-e^2\mu^{2\ve}\!\!\!\intfsD
  \bigg(
  \frac{p^2+m^2+2p{\cdot}s}{(s^2-m^2)(s-p)^2}+\xi\frac{1}{s^2}-(\xi-1)\frac{p^2-m^2}{s^2s^2}  \bigg)\,.
\end{equation}
The important point to note here is that, as with the seagull,  the quadratically divergent term is purely ultraviolet in nature, and can thus be robustly set to zero within dimensional regularisation. But the final, logarithmic divergence, tadpole is more subtle as it diverges in both the ultraviolet (regularised by taking $\ve>0$) and infrared (regularised by taking $\ve<0$) regimes. Extracting its ultraviolet divergence now leads to a non-vanishing, gauge dependent  contribution to the self-energy for the scalar particle. The end result is that the self-energy term (\ref{eq:scalar_se})  has a gauge dependent ultraviolet pole contribution, $-i\sigmaseuv$, which can be written in terms of the mass and inverse propagator as
\begin{equation}\label{eq:sigma_scalar}
  -i\sigmaseuv =-i\frac{e^2}{(4\pi)^2}\big(3m^2-i(\xi-3)\Props{m^2}^{-1}\big)\frac1{\varepsilon}\,.
 \end{equation} 

\bigskip

The vertex correction to the lowest order scalar interaction  with the background, (\ref{eq:v_int}), is then given by 
\begin{align}\label{eq:scalar_v_loop}
\begin{split}
  \raisebox{-0.6cm}{\includegraphics{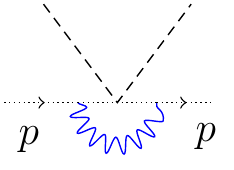}}=\Props{m^2}\big(-i\sigmama\big)\Props{m^2}\,,
\end{split}
\end{align}
where now the loop correction to the induced mass is
\begin{align}\label{eq:v_loop} 
\begin{split}
  -i\sigmama=-e^2\smstar^2\mu^{2\ve}\!\!\!\intfsD
  &\bigg(\frac{(p+s)^2}{(s^2-m^2)^2(s-p)^2}\\&+(\xi-1)\frac{(p^2-s^2)(p^2-s^2)}{(s^2-m^2)^2(s-p)^2(s-p)^2}\bigg)\,.
\end{split}
\end{align}  
The extra scalar propagator in the loop here ameliorates the divergences seen in the related self-energy (\ref{eq:scalar_sigma_se}), so that we now only have logarithmic terms to deal with. Thus, by naive power counting, both terms in (\ref{eq:v_loop}) will contribute the same ultraviolet pole, but the second term will have an additional multiplicative factor of  $\xi-1$. When combined we see  that the overall ultraviolet pole is proportional to the gauge fixing parameter $\xi$. Thus, writing this ultraviolet, double pole correction as $-i\sigmamauv$, we have at one-loop the gauge dependent result that 
\begin{equation}\label{eq:sigma_v}
  -i\sigmamauv=-i\frac{e^2}{(4\pi)^2}\xi \smstar^2\frac1{\varepsilon}\,.
\end{equation}
 
\bigskip

The one-loop ultraviolet results (\ref{eq:sigma_scalar}) and (\ref{eq:sigma_v}), and the expansion of the double line propagator (\ref{eq:scalar_doubleline}), are the only ingredients needed for building the full,  one-loop, ultraviolet corrections to the scalar propagator in the background.  This claim might seem surprising as we have only considered a single seagull interaction, but we note  that if a loop straddles more than one background interaction, then it is ultraviolet finite by simple power counting. This follows immediately from the extension of (\ref{eq:v_loop}) to that situation, where the power of $(s^2-m^2)$ in the denominator will then be  greater than two. This means that we only need to consider loops spanning single background seagulls in order to extract the ultraviolet terms. 

An immediate consequence of this, single seagull within a loop, result is a simple inductive characterisation of the loop corrections to multiple seagulls. 
We thus have the ultraviolet, loop factorisation identity that, for $r\ge0$,
\begin{equation}\label{eq:scalar_factorisation}
  \raisebox{-0.6cm}{\includegraphics{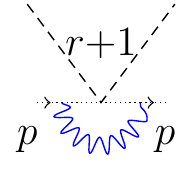}}\euv
  \raisebox{-0.6cm}{\includegraphics{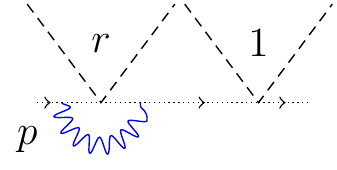}}+
  \raisebox{-0.6cm}{\includegraphics{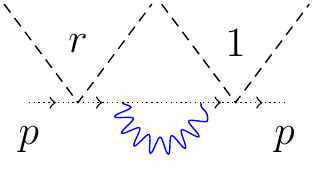}}+
  \raisebox{-0.6cm}{\includegraphics{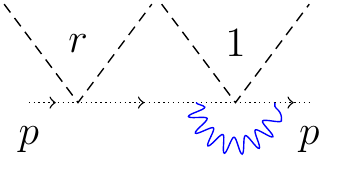}}\\\,.
\end{equation}
From this it follows by simple induction that if we define $\Delta(r)$, for $r\ge1$, by
\begin{equation}
  \Props{m^2}\deltauv(r)\euv
  \raisebox{-0.6cm}{\includegraphics{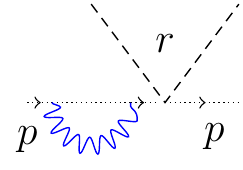}}+
  \raisebox{-0.6cm}{\includegraphics{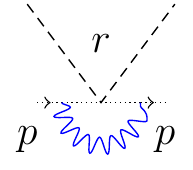}}+
  \raisebox{-0.6cm}{\includegraphics{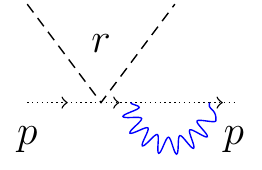}}\,,
\end{equation}
then
\begin{align}\label{eq:scalar_Delta}
\begin{split}
 \deltauv(r)=(r+1)&(-i\smstar^2\Props{m^2})^r(-i\sigmaseuv\Props{m^2})\\
  &+r(-i\smstar^2\Props{m^2})^{r-1}(-i\sigmamauv\Props{m^2})\,.
  \end{split}
\end{align}
Writing $\deltauv(0)=-i\sigmaseuv\Props{m^2}$, we then see that summing over all such degenerate processes yield the strong field, one-loop corrections 
\begin{align}\label{eq:scalar_allorders}
  \Props{m^2}\sum_{r=0}^\infty\deltauv(r)
  &=\Props{m^2}\sum_{r=1}^\infty r(-i\smstar^2\Props{m^2})^{r-1}\big(-i(\sigmaseuv+\sigmamauv)\big)\Props{m^2}\nonumber\\
  &=\Big(\Props{m^2}\sum_{j=0}^\infty (-i\smstar^2\Props{m^2})^{j}\Big)^2\big(-i(\sigmaseuv+\sigmamauv)\big)\,.
\end{align}
Recognising  in this expression the double line expansions (\ref{eq:scalar_doubleline}) squared, allows us to write (\ref{eq:scalar_allorders}) as the double line self-energy:
\begin{equation}\label{eq:scalar_doubleline_loop}
 \raisebox{-0.36cm}{\includegraphics{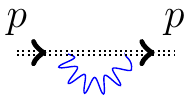}} \euv\Props{m^2+\smstar^2}\big(-i(\sigmaseuv+\sigmamauv)\big)\Props{m^2+\smstar^2}\,.
\end{equation}

This last result is a very succinct and attractive  summary of the ultraviolet, one-loop structure of the scalar QED propagator in our background. Although it was built up perturbatively in the background interactions, it is an all orders result and thus valid for both weak and strong background fields. Through the use of the momentum gauge to remove irrelevant clutter, we have recovered a very simple result that clearly identifies the ultraviolet divergences that need renormalising in this scalar theory. Indeed, we see here  a direct and simple link between the double line representation of a  loop contribution and the precise algebraic structure of the corresponding Green function. 
 
\bigskip

Having obtained this compact result for the scalar theory in the background which is structurally identical to that found  in a vacuum, we can now introduce counterterms   and renormalise using the familiar techniques of QED. But, before doing so, we shall first analyse  what happens with fermionic matter. We will then return to discuss the renormalisation of both theories in section~\ref{sec:renormalisation}.

\section{Fermionic matter}
The impact of the plane wave background on fermionic matter includes a mass shift to $m_*^2$, first identified in~\cite{Volkov:1935zz} and \cite{Sengupta:1952}, which was later seen to have a more subtle, matrix structure in~\cite{Lavelle:2015jxa}. The background also generates  an infinite class of sidebands that permeate the theory \cite{Reiss:1966A}\cite{Eberly:1966b}\cite{Lavelle:2017dzx}\cite{Lavelle:2019lys}, resulting in momentum shifts and additional spacetime phases. 
We will see, though, that the use of the momentum gauge streamlines the route to the induced mass and also reduces the  sidebands to a finite number.

To lay the groundwork for these results, and their one-loop extensions, we will first,  in this section, introduce the matrix structures associated with the fermions and identify the key  simplifications that follow from the use of the momentum gauge and our choice of polarisation.  We will then apply these results in the following sections to both the tree and one-loop description of the full propagator in our background. 

\medskip
 
The absorption by an electron of a  photon from the background    is now characterised by the absorption matrix, $\Ab$, which is given in terms of  the complex potential, $\laser^\mu$, by 
\begin{equation}\label{eq:Ab}
  \Ab=-i\lasersl\,.
\end{equation}
The dual matrix, $\Em$, describing the emission of a photon degenerate to the background, is then given by 
\begin{equation}\label{eq:Em}
  \Em=-i\laserssl\,.
\end{equation} 
These are the fermionic counterparts to the scalar terms (\ref{eq:s_ab}) and (\ref{eq:s_em}), but they do not now vanish in the momentum gauge. Indeed, these are now the only interactions with the background for the fermion, as there is no equivalent to the seagull term that was central to the scalar theory.

In terms of these absorption  and emission  matrices, the gauge conditions (\ref{eq:null_gauge}) and (\ref{eq:mmt_gauge}) imply the anti-commutation results that
\begin{equation}\label{eq:k_comm}
  \ksl\Ab=-\Ab\ksl\,,\qquad \ksl\Em=-\Em\ksl\,,
\end{equation}
and
\begin{equation}\label{eq:p_comm}
  \psl\Ab=-\Ab\psl\,,\qquad \psl\Em=-\Em\psl\,.
\end{equation} 

Just as we did for the scalar field, it is useful to introduce a compact notation for the fermionic propagator, but now it needs to incorporate the  degeneracy induced by the background that was alluded to earlier. We thus define, for integer $n$, the shifted fermionic propagator, $\Prop{(m)}{n}$,  by
\begin{equation}\label{eq:Prop}
  \Prop{(m)}{n}=\frac{i}{\psl+n\ksl-m+i\epsilon}\,.
\end{equation}
Diagrammatically, these fermionic propagators will be represented by the usual plane line. Thus the fundamental absorption process from the background is now given by 
\begin{equation}\label{eq:fermi_ab}
 \raisebox{-0.52cm}{\includegraphics{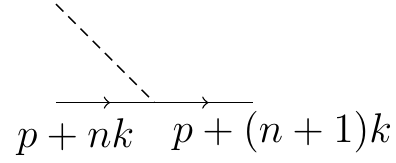}} = \Prop{(m)}{n+1}\big(-i\lasersl\big)\Prop{(m)}{n}\equiv \Prop{(m)}{n+1}\Ab\Prop{(m)}{n}\,,
\end{equation} 
while the emission process to the background is
\begin{equation}\label{eq:fermi_em}
 \raisebox{-0.52cm}{\includegraphics{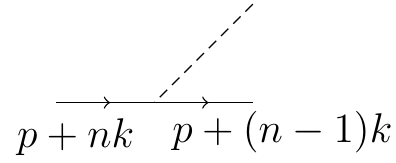}} = \Prop{(m)}{n-1}\big(-i\laserssl\big)\Prop{(m)}{n}\equiv \Prop{(m)}{n-1}\Em\Prop{(m)}{n}\,. 
\end{equation}

The linear mass dependence and the extra subscript in (\ref{eq:Prop}) will help to distinguish this propagator from the scalar one, $\Props{m^2}$, introduced in (\ref{eq:props}). Obviously, but very importantly, the big technical difference between $\Prop{(m)}{n}$ and $\Props{m^2}$ that must be kept in mind  is that the fermionic one is a matrix. 

The algebraic complexities of the fermionic theory mean that it will often be useful to abbreviate this fermionic propagator by suppressing the explicit  mass term, as in  $\Prop{(m)}{n}\to\Prop{}{n}$. Indeed, we will use this condensed notation for the fermionic propagator in what follows and only reintroduce the more explicit  form after equation (\ref{eq:mass_shift2}).

\bigskip

From our discussion of the scalar theory, we know that the propagator $\Prop{}{0}$ describes both the usual  free propagator of momentum $p$  and also the propagator  where the number of absorptions equals the number of emissions.  Obviously, a similar degeneracy will arise for each propagator, $\Prop{}{n}$. But, in addition, a mismatch between the number of emissions and absorptions will result in a shift in the value of $n$. This process is described using the interactions (\ref{eq:Ab}) and (\ref{eq:Em}) in the standard way, by considering the vertex term  $\Prop{}{n+1}\Ab\Prop{}{n}$ for an absorption and its dual $\Prop{}{n}\Em\Prop{}{n+1}$ for an emission. However, what is not standard from a field theory point of view is the fact that these interactions can be rewritten as the difference of two distinct propagators, and that this holds at all orders in the background interactions. This Ward type of identity  leads to the sideband description of the charge   that was first described in \cite{Reiss:1966A} and further refined in \cite{Lavelle:2019lys}.

In a perturbative framework, the emergence of sidebands is simply a partial fraction expansion of the absorption, (\ref{eq:fermi_ab}), and emission, (\ref{eq:fermi_em}), interactions. This quickly leads to  the  key absorption and emission identities that
\begin{equation}\label{eq:SbP}
\Prop{}{n+1}\Ab\Prop{}{n}=\In\Prop{}{n}-\Prop{}{n+1}\In\qquad\mathrm{and}\qquad\Prop{}{n-1}\Em\Prop{}{n}=\Prop{}{n-1}\Out-\Out\Prop{}{n}\,.
\end{equation} 
The existence of this partial fraction decomposition only relies on the light cone  property of $k^\mu$ and the null gauge condition (\ref{eq:null_gauge}). But the form of the \lq In\rq\ factor, $\In$, and the \lq Out\rq\ factor, $\Out$, is sensitive to the momentum gauge choice,~(\ref{eq:mmt_gauge}). We now find that
\begin{equation}\label{eq:In}
  \In:=\frac{2p\cdot\laser+\ksl\lasersl}{2\pk}=\frac{\ksl\lasersl}{2\pk}\,,
\end{equation}
while its dual \lq Out\rq\ matrix is
\begin{equation}\label{eq:Out}
  \Out:=\frac{2p\cdot\lasers+\laserssl\ksl}{2\pk}=\frac{\laserssl\ksl}{2\pk}\,.
\end{equation}

We have now introduced all the basic variables needed to build up a description of the electron propagating through the background. The way to proceed, that makes the transition to the loop corrections most transparent, is to now shift from the language of absorptions, (\ref{eq:Ab}), and emissions, (\ref{eq:Em}), to that of the \lq In\rq, (\ref{eq:In}), and \lq Out\rq, (\ref{eq:Out}), factors. This perturbative refocussing of the formalism for fermions will be the topic of the next section.  Prior to embarking on that, though, it is useful to conclude this section with a summary of the key new simplifications that follow  from our choice of circular polarisation, (\ref{eq:A_circular}),  and the additional momentum gauge condition, (\ref{eq:mmt_gauge}).

The immediate impact of using a circular polarisation is that polynomials in the interactions become trivial. In particular, we have already noted that from~(\ref{eq:A_circular}), $\laser\cdot\laser=\lasers\cdot\lasers=0$. This means that the scalar term $v$ and $v^*$ that play an important role in the full elliptical class of polarisation, see \cite{Lavelle:2017dzx}, now vanish:
\begin{equation}\label{eq:v_vanish}
 v:=\frac{\laser{\cdot}\laser}{2\pk}=0\qquad \mathrm{and}\quad v^*:=\frac{\lasers{\cdot}\lasers}{2\pk}=0\,.
\end{equation} 
In terms of the absorption  and emission matrices, these polarisation dependent simplifications become
\begin{equation}\label{eq:simp_AandE}
 \Ab^2=0\qquad \mathrm{and}\quad \Em^2=0\,.
\end{equation}
These last two results are easily extended using the momentum gauge conditions~(\ref{eq:p_comm}), so that 
\begin{equation}\label{eq:simp_AandEwithp}
 \Ab\psl\Ab=0\quad \mathrm{and}\quad \Em\psl\Em=0\,.
\end{equation}
Written in terms of the propagators (\ref{eq:Prop}), these last identities become 
\begin{equation}\label{eq:simp_AandEwithP}
 \Ab\Prop{}{n}\Ab=0\quad \mathrm{and}\quad \Em\Prop{}{n}\Em=0\,.
\end{equation}

For the  \lq In\rq\ and  \lq Out\rq\ factors we have similar algebraic properties to (\ref{eq:simp_AandE}), but now these are polarisation independent and just follow from the momentum gauge choice, so that
\begin{equation}\label{eq:simp_ininoutout}
 \In^2=0\quad\mathrm{and}\quad \Out^2=0\,.
\end{equation} 
In addition, the momentum gauge choice also implies the polarisation independent, trivial mixed products of these factors: 
\begin{equation}\label{eq:simp_inout}
  \In\Out=0\quad\mathrm{and}\quad \Out\In=0\,.
\end{equation} 

The choice of circular polarisation combines with the momentum gauge to now give momentum insertion identities similar to (\ref{eq:simp_AandEwithp}), namely $\In\psl\In=0$ and $\Out\psl\Out=0$. 
These simple results have two very important refinements that will be repeatedly used and extended in our analysis. If we first view the momentum term here as coming from the propagator (\ref{eq:Prop}), then we have 
\begin{equation}\label{eq:simp_Propp}
  \In\Prop{}{n}\In=0\quad \mathrm{and}\quad \Out\Prop{}{n}\Out=0\,.
\end{equation} 
While if we identify the momentum as part of the inverse propagator \\ \hbox{$\Prop{}{n}^{-1}=-i(\psl+n\ksl-m)$}, then we have 
\begin{equation}\label{eq:simp_IProp}
  \In\Prop{}{n}^{-1}\In=0\quad \mathrm{and}\quad \Out\Prop{}{n}^{-1}\Out=0\,.
\end{equation}

It is important to note that the mixed identities in (\ref{eq:simp_inout}) are not reflected in the product of an absorption and an emission matrix. Indeed, from (\ref{eq:A_circular}), we have \hbox{$\lasers\cdot\laser=\frac12 e^2 a^2$}, which does not vanish. This result is  actually polarisation independent, see \cite{Lavelle:2017dzx}. In terms of the absorption and emission matrices, we write this last key identity as
\begin{equation}\label{eq:mass_id}
  \Ab\Em+\Em\Ab=-2\lasers\cdot\laser=2p\cdot\mstar\,,
\end{equation} 
where the oxymoronic,  \lq mass null vector\rq\ is defined by
\begin{equation}\label{eq:null_mass}
  \mstar^\mu= -\frac{\lasers{\cdot}\laser}{p\cdot k}k^\mu\,.
\end{equation}
Note that from the scalar mass definition (\ref{eq:mstar2}), we have the vector mass identity that
\begin{equation}
  \smstar^2=2p\cdot\mstar\,.
\end{equation}

In terms of the \lq In\rq\ and \lq Out\rq\ factors, the null vector mass term arises from the momentum insertions whereby $\In\psl\Out+\Out\psl\In = \mstarsl$. Written in terms of the inverse propagator, this becomes
\begin{equation}\label{eq:in_iprop_out}
  \In\Prop{}{n}^{-1}\Out+\Out\Prop{}{n}^{-1}\In = -i\mstarsl\,.
\end{equation}
Finally, we note that the propagator identities in (\ref{eq:simp_Propp}) have the immediate mass insertion generalisations that, for $r\ge0$,
\begin{equation}\label{eq:in_Propntor_in}
  \In\Prop{}{n}(-i\mstarsl\Prop{}{n})^r\In=0\quad \mathrm{and}\quad \Out\Prop{}{n}(-i\mstarsl\Prop{}{n})^r\Out=0\,.
\end{equation}

\section{Tree level propagation} Loop corrections to the propagation of the charge are most readily   introduced via a perturbative formulation of the tree level results. We shall now develop such a description, taking full advantage of the simplifications that follow from using the momentum gauge.  Armed with these results, we shall then be ready to add one-loop corrections to these all orders interactions with the background.

\bigskip
Building upon our definition of the  scalar double line propagator, (\ref{eq:scalar_doubleline}), we define the fermionic double line propagator, of momentum $p$, to be the sum over all possible tree level, perturbative interactions with the background that start with momentum~$p$,
\begin{equation}\label{eq:many_in_out}
\raisebox{-0.08cm}{\includegraphics{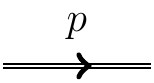}}=\sum_{r_1=0}^{\infty}\sum_{r_2=0}^{\infty} 
\raisebox{-0.5cm}{\includegraphics{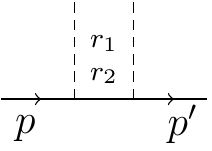}} \,.
\end{equation}
The notation being introduced here is the generalisation of the scalar result (\ref{eq:scalar_doubleline}) to the situation where we have both absorptions and emissions from the background. So the sum is now over all processes with $r_1$ absorptions and $r_2$ emissions, degenerate to the background. In contrast to the scalar theory,  each such process will in general correspond to multiple Feynman diagrams. Note that momentum conservation implies that the out going momentum is \hbox{$p'=p+(r_1-r_2)k$}.

The single absorption process, (\ref{eq:fermi_ab}), but in which the charge with initial momentum $p$ absorbs a background photon, is given in terms of the propagator,  (\ref{eq:Prop}), by the vertex contribution $\Prop{}{1}\Ab\Prop{}{0}$. Thus we have, from (\ref{eq:SbP}), the sideband representation of this interaction:
\begin{equation}\label{eq:1_in_0_out} 
\raisebox{-0.55cm}{\includegraphics{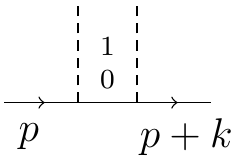}}=\In\Prop{}{0}-\Prop{}{1}\In\,.
\end{equation}
In a similar way,  the single emission process, (\ref{eq:fermi_em}),  becomes
\begin{equation}\label{eq:0_in_1_out} 
\raisebox{-0.55cm}{\includegraphics{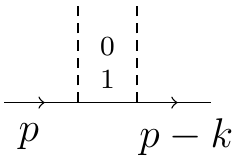}}=\Prop{}{-1}\Out-\Out\Prop{}{0}\,.
\end{equation} 
If we now consider two absorptions, then we get from (\ref{eq:simp_AandEwithP}) the vanishing result  that
\begin{equation}\label{eq:2_in_0_out} 
\raisebox{-0.55cm}{\includegraphics{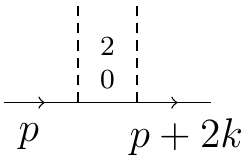}}=\Prop{}{2}\Ab\Prop{}{1}\Ab\Prop{}{0}=0\,. 
\end{equation}

This last example can be easily generalised so that if the difference between the number of absorptions and emissions is greater than one, then the contribution to the double line propagator, (\ref{eq:many_in_out}), vanishes:
\begin{equation}\label{eq:n1_in_n2_out_zero}
\raisebox{-0.45cm}{\includegraphics{fermion_many_in_out.pdf}}=0\quad \mathrm{if}\quad |r_1-r_2|>1\,.
\end{equation}  
This key vanishing result follows from both our choice of momentum gauge and polarisation. The proof is straightforward  since all perturbative  contributions in (\ref{eq:n1_in_n2_out_zero}) must now include parts where we have either two consecutive absorptions or two consecutive emissions, separated by an appropriate  propagator $\Prop{}{n}$. These then vanish by the identities (\ref{eq:simp_AandEwithP}).

From the vanishing result~(\ref{eq:n1_in_n2_out_zero}), we see that the only  other non-vanishing contributions to the propagator (\ref{eq:many_in_out})  arise when the absorptions alternate with the emissions. The lowest order terms of this form are given by the the processes whereby  the electron first absorbs a photon and then emits back into the background, or first emits then absorbs from the background: $\Prop{}{0}\big(\Ab\Prop{}{-1}\Em+\Em\Prop{}{1}\Ab\big)\Prop{}{0}$. The central factor here can be seen as a propagator insertion into the mass term~(\ref{eq:mass_id}). One quickly finds that the momentum gauge implies the polarisation independent result that
\begin{equation}
  \Ab\Prop{}{-1}\Em+\Em\Prop{}{1}\Ab=-i\mstarsl+\Prop{}{0}^{-1}\big(\In\Prop{}{-1}\Out+\Out\Prop{}{1}\In\big)\Prop{}{0}^{-1}\,.
\end{equation}
Hence we see that
\begin{equation}\label{eq:1_in_1_out} 
\raisebox{-0.5cm}{\includegraphics{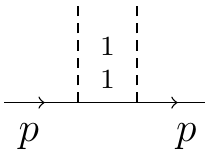}}=\In\Prop{}{-1}\Out+\Prop{}{0}(-i\mstarsl)\Prop{}{0}+\Out\Prop{}{1}\In\,.
\end{equation} 
This is the fermionic version of the scalar seagull term (\ref{eq:v_int}). Again we note that the fermionic theory has even sidebands associated with this central term. 
  
 \bigskip

The lowest order results (\ref{eq:1_in_0_out}), (\ref{eq:0_in_1_out}) and (\ref{eq:1_in_1_out}) can now be extended to all orders in the background interaction. Key to that extension is the following factorisation result, that is derived in Appendix~\ref{sec:appendix1}, valid for $r_1\ge0$ and $r_2\ge0$:
\begin{equation}\label{eq:factorise} 
\raisebox{-0.55cm}{\includegraphics{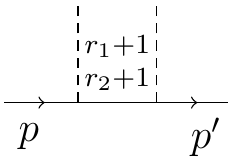}}=
\raisebox{-0.55cm}{\includegraphics{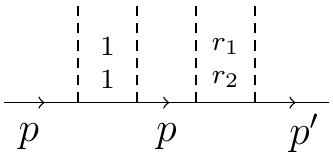}}=
\raisebox{-0.55cm}{\includegraphics{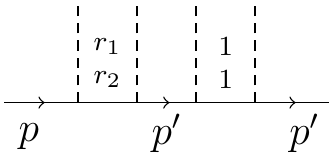}}
\,.
\end{equation}  
Given the mass generating term in (\ref{eq:1_in_1_out}), the factorisation in (\ref{eq:factorise}) will spawn background induced mass terms into all the sidebands seen in (\ref{eq:1_in_0_out}), (\ref{eq:0_in_1_out}) and (\ref{eq:1_in_1_out}). 
Exploiting this factorisation then quickly leads to the results, also discussed  in Appendix~\ref{sec:appendix1}, that for all $r\ge0$:
\begin{equation}\label{eq:ind1}
\raisebox{-0.5cm}{\includegraphics{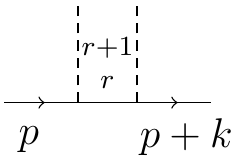}} =\In\Prop{}{0}\big(-i\mstarsl\Prop{}{0}\big)^{r}-\big(\Prop{}{1}(-i\mstarsl)\big)^{r}\Prop{}{1}\In\,,
\end{equation}   
\begin{equation}\label{eq:ind2}
\raisebox{-0.5cm}{\includegraphics{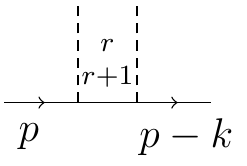}} =\big(\Prop{}{-1}(-i\mstarsl)\big)^{r}\Prop{}{-1}\Out-\Out\Prop{}{0}\big(-i\mstarsl\Prop{}{0}\big)^{r}
\end{equation} 
and 
\begin{equation}\label{eq:ind3}
\raisebox{-0.5cm}{\includegraphics{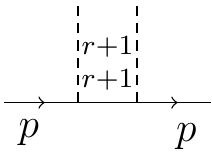}}=\In\Prop{}{-1}\big(-i\mstarsl\Prop{}{-1}\big)^{r}\Out+\Prop{}{0}\big(-i\mstarsl\Prop{}{0}\big)^{r+1}+\Out\Prop{}{1}\big(-i\mstarsl\Prop{}{1}\big)^{r}\In\,.
\end{equation}

Using the vanishing result (\ref{eq:n1_in_n2_out_zero}) in the double line definition (\ref{eq:many_in_out}), we see that the double sum over interactions becomes the single sum:
\begin{equation}\label{eq:doubleline1}
  \raisebox{-0.1cm}{\includegraphics{fermion_doubleline.pdf}} =\sum_{r=0}^{\infty} 
  \raisebox{-0.55cm}{\includegraphics{fermion_r_plus1_in_r_out.pdf}}+
  \raisebox{-0.55cm}{\includegraphics{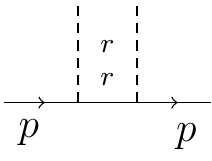}}+
  \raisebox{-0.55cm}{\includegraphics{fermion_r_in_r_plus1_out.pdf}}.
\end{equation}  
The terms being summed over here are explicitly given by the previous key results  (\ref{eq:ind1}), (\ref{eq:ind2}) and (\ref{eq:ind3}).   

To interpret this representation of the double line propagator for fermionic matter, it is helpful to reinstate the explicit mass dependence  of the propagator, so that  $\Prop{}{n}\to\Prop{(m)}{n}$. Mimicking  the scalar argument in (\ref{eq:scalar_mass_shift}), if the fermionic mass $m$ is now shifted by the matrix term $\mstarsl$,  then we get the expansion
\begin{align}\label{eq:mass_shift2}
  \Prop{(m+\mstarsl)}{n}&=\frac{i}{\psl+n\ksl-(m+\mstarsl)+i\epsilon}\nonumber\\
  &=\sum_{r=0}^{\infty} \Big(\Prop{(m)}{n}(-i\mstarsl)\Big)^r\Prop{(m)}{n}\,,
\end{align}
where in (\ref{eq:mass_shift2}) the single propagator term can also be factored out to the left. 

Using this mass shift identity allows us to rewrite  in a very succinct way the all orders tree level result  (\ref{eq:doubleline1}) as the core  sideband expansion:
\begin{align}\label{eq:doubleline2}
\raisebox{-0.05cm}{\includegraphics{fermion_doubleline.pdf}}&= 
\In\Prop{(m+\mstarsl)}{0}-\Prop{(m+\mstarsl)}{1}\In\nonumber\\ &\quad+\In\Prop{(m+\mstarsl)}{-1}\Out+\Prop{(m+\mstarsl)}{0}+\Out\Prop{(m+\mstarsl)}{1}\In \\ &\qquad+\Prop{(m+\mstarsl)}{-1}\Out-\Out\Prop{(m+\mstarsl)}{0}. \nonumber
\end{align}
Note that in this all orders, tree level expression for the fermionic double line propagator,  the  upper terms  have spacetime dependence inherited from the \lq In\rq\ factor of $\ee^{-i\xk}$, the  central terms have no such spacetime factors, while the lower terms  inherited from the \lq Out\rq\ factor an $\ee^{i\xk}$ dependence.  

It is also useful to note that the fermionic propagator $\Prop{(m+\mstarsl)}{n}$ in these last few expressions  can also  be partially written in terms of the scalar mass $\smstar$, introduced in (\ref{eq:mstar2}), as
\begin{equation}
  \Prop{(m+\mstarsl)}{n}=\frac{i(\psl+n\ksl+m-\mstarsl)}{(p+nk)^2-(m^2+\smstar^2)+i\epsilon}\,.
\end{equation}
This representation makes clear the new pole structure in the sidebands and highlights the fundamental  difference in this fermionic theory due to the vector nature of the induced mass term in the numerator.

The expression (\ref{eq:doubleline2}) for the all-orders, tree level, fermionic  propagator in the background is surprisingly compact, with a very manageable number of core sidebands. In Appendix~\ref{sec:app_ritus} we show how this formulation of the fermionic double line propagator relates to the more standard discussions found in the literature.

\section{One-loop correction in a weak background} 
Having constructed the tree level, double line fermionic propagator in (\ref{eq:doubleline1}) and (\ref{eq:doubleline2}), we now want to incorporate into these sideband expressions their  one-loop corrections. Just as for the scalar theory, this will be built up perturbatively  over the interactions with the background. So we will start our analysis in this section by considering a weak background and hence looking at the loop corrections to the lowest order background  terms introduced earlier: the single absorption (\ref{eq:1_in_0_out}), the single emission (\ref{eq:0_in_1_out}), and both processes with a single absorption and emission~(\ref{eq:1_in_1_out}). These lowest order one-loop corrections were first discussed in \cite{Lavelle:2019lys}, and extended to the full Lorentz class of gauges in \cite{Lavelle:2019vcz}.  Now we shall exploit the simplifications  that arise due to our momentum gauge choice, (\ref{eq:mmt_gauge}), to give a   more direct account of these results for a circularly polarised background.   
  
\bigskip 
Using a simple notational extension of the scalar self-energy term introduced in (\ref{eq:scalar_se}), but now allowing for a sideband momenta of $p+nk$, we take the fermionic self-energy to be given by the expression
\begin{equation}\label{eq:fermion_se}
  \raisebox{-0.6cm}{\includegraphics{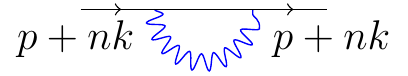}}=\Prop{}{n}\big(-i\sigmafse(n)\big)\Prop{}{n}\,.
\end{equation} 
Note, though, that by translational invariance, we can focus on the structure of the central sideband here, with $n=0$, and then replace $p$ by $p+nk$ for the more general result in what follows. 

We thus have, within our Lorentz class of gauge fixings for the loop,
\begin{align}\label{eq:fermion_sigma_se}
\begin{split}
  -i\sigmafse(0)=-e^2\mu^{2\ve}\!\!\!\intfsD
  &\bigg(\frac{\gamma^\mu(\ssl+m)\gamma_\mu}{(s^2-m^2)(s-p)^2}\\&+(\xi-1)\frac{(\ssl-\psl)(\ssl+m)(\ssl-\psl)}{(s-p)^2(s^2-m^2)(s-p)^2}\bigg)\,.
\end{split}
\end{align}
In contrast to the equivalent scalar version, (\ref{eq:scalar_sigma_se}), there are now no quadratic ultraviolet divergences, but there are still linear and logarithmic ones to be identified. Simple power counting arguments now quickly show that 
\begin{equation}
   -i\sigmafse(0)\euv-e^2\!\!\!\intfsD\Big( (3m-(\psl-m))\frac1{s^2s^2}-(\xi-1)(\psl-m)\frac1{s^2s^2}\Big)\,.
\end{equation}
Thus the fermionic version of the ultraviolet scalar result (\ref{eq:sigma_scalar}), adapted to the $n^\mathrm{th}$ sideband, is the self-energy expression that
\begin{equation}\label{eq:sigma_uvse_fermion}
  -i\sigmafseuv(n) =-i\frac{e^2}{(4\pi)^2}\big(3m-i\xi\Prop{}{n}^{-1}\big)\frac1{\varepsilon}\,.
 \end{equation} 

An analysis of the vertex corrections in the fermionic theory is more involved than in the scalar theory, due to the associated changes in sideband structures related to  whether we have an absorption or an emission or both. To unpick this we start with the vertex correction to the fundamental absorption process (\ref{eq:1_in_0_out}).

\medskip

The  vertex correction to the absorption of an in-coming background photon is given by
\begin{equation}\label{eq:fermion_in_loop}
  \raisebox{-0.42cm}{\includegraphics{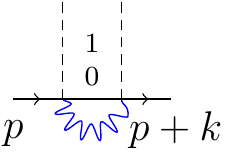}}:=\Prop{}{1}\big(-i\sigmain\big)\Prop{}{0}\,,
\end{equation}
where
\begin{align}\label{eq:fermion_sigma_in}
\begin{split}
  -i\sigmain=-e^2\mu^{2\ve}\!\!\!\intfsD
  &\bigg(\frac{\gamma^\mu(\ssl+\ksl+m)\lasersl(\ssl+m)\gamma_\mu}{((s+k)^2-m^2)(s^2-m^2)(s-p)^2}\\&+(\xi-1)\frac{(\ssl-\psl)(\ssl+\ksl+m)\lasersl(\ssl+m)(\ssl-\psl)}{(s-p)^2((s+k)^2-m^2)(s^2-m^2)(s-p)^2}\bigg)\,.
\end{split}
\end{align}
Again, simple power counting arguments quickly show that 
\begin{equation}
   -i\sigmain\euv-e^2\!\!\!\intfsD\Big( \lasersl\frac1{s^2s^2}-(\xi-1)\lasersl\frac1{s^2s^2}\Big)\,.
\end{equation}
Recalling our definition of the absorption matrix, (\ref{eq:Ab}), this quickly leads to the ultraviolet, in-coming, vertex contribution
\begin{equation}\label{eq:fermion_ab_uv}
   -i\sigmainuv:=\frac{e^2}{(4\pi)^2}\xi\Ab\frac1{\varepsilon}\,.
\end{equation}

Using the sideband representation of the absorption matrix, (\ref{eq:SbP}), allows us to rewrite this vertex contribution in terms of consecutive sideband self-energies, (\ref{eq:sigma_uvse_fermion}), so that 
\begin{equation}\label{eq:fermion_ab_as_se_uv}
  -i\sigmainuv=\In \big(-i\sigmafseuv(n)\big)-\big(-i\sigmafseuv(n+1)\big)\In\,.
\end{equation}

\medskip

To understand the significance of the one-loop results, (\ref{eq:sigma_uvse_fermion}) and (\ref{eq:fermion_ab_as_se_uv}), we now consider the full, one-loop corrections to the single absorption process as given by the two self-energy and one vertex corrections:
\begin{align}\label{eq:One_loop_10}
  \raisebox{-0.45cm}{\includegraphics{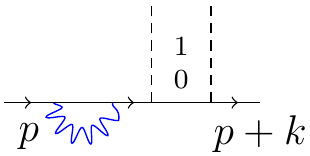}}&+\quad
  \raisebox{-0.45cm}{\includegraphics{fermion_laser_in_loop.pdf}}+\quad
  \raisebox{-0.45cm}{\includegraphics{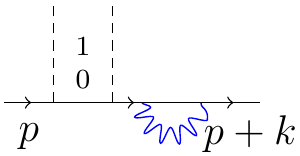}}\nonumber\\
  &\euv\big(\In\Prop{}{0}-\Prop{}{1}\In\big)(-i\sigmafseuv(0))\Prop{}{0}
  +\Prop{}{1}(-i\sigmainuv)\Prop{}{0}\nonumber\\
  &\qquad\qquad+\Prop{}{1}(-i\sigmafseuv(1))\big(\In\Prop{}{0}-\Prop{}{1}\In\big)\nonumber\\&=\In\Prop{}{0}(-i\sigmafseuv(0))\Prop{}{0}-\Prop{}{1}(-i\sigmafseuv(1))\Prop{}{1}\In\,.
\end{align}
This we recognise as the naively expected, one loop self-energy corrections to the sidebands in (\ref{eq:1_in_0_out}).

\medskip

In a very similar way, the leading one-loop vertex correction to the  dual emission process, (\ref{eq:0_in_1_out}), is given by  
\begin{equation}
  \raisebox{-0.45cm}{\includegraphics{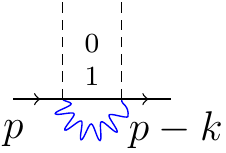}} \euv  \Prop{}{-1}\big(-i\sigmaoutuv\big)\Prop{}{0}\,,
\end{equation} 
where the out-going version of the in-coming vertex, (\ref{eq:fermion_ab_uv}), and its sideband representation (\ref{eq:fermion_ab_as_se_uv}),  are now
\begin{equation}\label{eq:out_loop}
-i\sigmaoutuv=\frac{e^2}{(4\pi)^2}\xi\Em\frac1{\varepsilon}=\big(-i\sigmafseuv(n-1)\big)\Out-\Out\big(-i\sigmafseuv(n)\big)\,.
\end{equation}
Hence we quickly see that the 
one-loop corrections to the single emission process are:
\begin{align}\label{eq:One_loop_01}
  \raisebox{-0.45cm}{\includegraphics{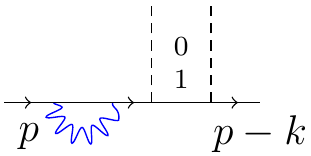}}&+
  \raisebox{-0.45cm}{\includegraphics{fermion_laser_out_loop.pdf}}+
  \raisebox{-0.45cm}{\includegraphics{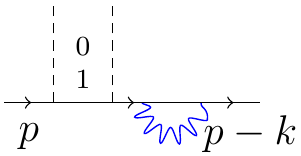}}\nonumber\\& \euv  \Prop{}{-1}(-i\sigmafseuv(-1))\Prop{}{-1}\Out-\Out\Prop{}{0}(-i\sigmafseuv(0))\Prop{}{0}\,.
\end{align}
This is also the  expected, one-loop self-energy corrections to the absorption sidebands in (\ref{eq:0_in_1_out}).

\bigskip

Loop corrections to processes with a mixture of emissions and absorptions introduce an additional, but gauge dependent, ultraviolet divergence associated with the background induced mass term, (\ref{eq:null_mass}). This was first identified at lowest order in the background,  for Feynman gauge, in \cite{Lavelle:2019lys} and then extended to the full Lorentz class in \cite{Lavelle:2019vcz}. We have already seen here, in expression (\ref{eq:1_in_1_out}), that the identification of the background induced mass is simplified by the use of the momentum gauge. Now we will see how that gauge also streamlines  the discussion of this new  ultraviolet correction.

\medskip

The one-loop corrections to the lowest order mixed absorption and emission process,~(\ref{eq:1_in_1_out}), can clearly spawn simple self-energy terms. But these corrections can  also straddle more complex, interaction structures associated with the background. Indeed, the  vertex term here is now a mixture of emissions and absorptions, and the loop corrections are thus more involved. Focusing, though, on the ultraviolet structure leads to a simple factorisation of these vertex corrections, the  details of which are discussed  in Appendix~\ref{sec:app_mass}. One finds 
\begin{align}\label{eq:1_1_loop}
  \raisebox{-0.42cm}{\includegraphics{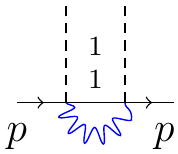}}\euv\In&\Prop{}{-1}(-i\sigmafseuv(-1))\Prop{}{-1}\Out+\Prop{}{0}\big(-i\sigmasluv\big)\Prop{}{0} \nonumber\\[-0.4cm]&\hspace{2cm}+\Out\Prop{}{1}(-i\sigmafseuv(1))\Prop{}{1}\In\\
  &-\Prop{}{0}(-i\sigmafseuv(0))\big(\In\Prop{}{-1}\Out+\Out\Prop{}{1}\In\big)
  -\big(\In\Prop{}{-1}\Out+\Out\Prop{}{1}\In\big)(-i\sigmafseuv(0))\Prop{}{0}\,.\nonumber
\end{align}
Combining this vertex contribution with the simpler self-energy corrections  gives the ultraviolet pole identification that 
\begin{align}\label{eq:One_loop_11}
\begin{split}
  \raisebox{-0.42cm}{\includegraphics{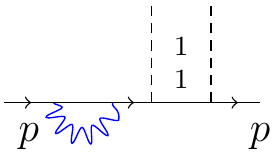}}&+
  \raisebox{-0.42cm}{\includegraphics{fermion_laser_line_loop_11.pdf}}+
  \raisebox{-0.42cm}{\includegraphics{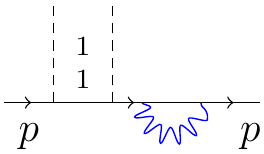}}\\ \euv\In\Prop{}{-1}&(-i\sigmafseuv(-1))\Prop{}{-1}\Out\\&\hspace{0.5cm}+\Prop{}{0} (-i\sigmafseuv(0))\Prop{}{0}\big(-i\mstarsl\big)\Prop{}{0}+\Prop{}{0}\big(-i\sigmasluv\big)\Prop{}{0} \\&\hspace{2cm}+\Prop{}{0}\big(-i\mstarsl\big)\Prop{}{0}(-i\sigmafseuv(0))\Prop{}{0} \\&\hspace{1cm}+\Out\Prop{}{1}(-i\sigmafseuv(1))\Prop{}{1}\In
  \end{split}
\end{align}
As seen earlier for the simpler absorption and emission processes,  this contains the expected propagator renormalisation terms for each sideband. But, in addition, the middle term in the central sideband here shows the gauge dependent, ultraviolet divergence related to the background induced mass, where:
\begin{equation}\label{eq:massUV}
  -i\sigmasluv=-i\frac{e^2}{(4\pi)^2}\xi\mstarsl\frac1\varepsilon
  \,.
\end{equation} 

The results presented in (\ref{eq:One_loop_10}), (\ref{eq:One_loop_01}) and (\ref{eq:One_loop_11}) are the lowest order, ultraviolet corrections to the fundamental interactions of the electron with a circularly polarised but weak background. They show a  sideband structure, which leads to a multiplicative renormalisation at this order, as discussed in \cite{Lavelle:2019lys} and \cite{Lavelle:2019vcz}. The use of the momentum gauge has greatly simplified this  analysis. In the next section we consider  the extension of these results to the strong field situation where results at all orders in the background field are needed.

\section{One loop correction in a strong background} \label{sec:strong_fermion_loop}
Extending the  weak field, one-loop results to strong field QED will follow  the basic steps seen earlier for scalar matter, but now this needs to be done within each of the sidebands. Thus the fermionic version of the scalar double line propagator, (\ref{eq:scalar_doubleline_loop}), will now involve an additional sum over all possible sidebands.  In this section we will  present the key steps in arriving at such a description of the fermion propagating through such an intense background.  Some of the important technical details of this account will be collected together in Appendix~\ref{sec:app_strong}. 

The ultraviolet factorisation, seen in (\ref{eq:scalar_factorisation}), was the key technical tool in our analysis of the one-loop results for scalar matter. We have already seen the tree-level version of this for fermionic matter in (\ref{eq:factorise}) and, in expression (\ref{eq:loop_factorise}) of Appendix~\ref{sec:app_strong}, we describe its one-loop version. Using it, we quickly see that the strong background interaction version of the absorption loop corrections, (\ref{eq:One_loop_10}), can be written as
\begin{align}\label{eq:One_loop_rp1r}
  \raisebox{-0.45cm}{\includegraphics{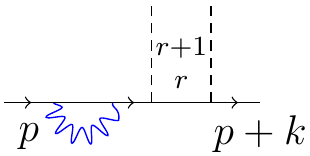}}&+\quad
  \raisebox{-0.45cm}{\includegraphics{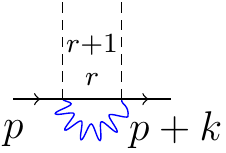}}+\quad
  \raisebox{-0.45cm}{\includegraphics{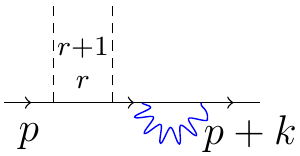}}\nonumber\\
  &\euv\In\Prop{}{0}\deltauv_{0}(r)-\Prop{}{1}\deltauv_1(r)\In\,,
\end{align}
for $r\ge0$. 
In this expression  we have extended to fermions, and their sidebands, the  notation, $\deltauv(r)$, introduced for scalar matter in (\ref{eq:scalar_Delta}). Now the $n^{\mathrm{th}}$ sideband at order $r$ has a factor $\deltauv_n(r)$ in (\ref{eq:One_loop_rp1r}) where, if $r\ge1$,
\begin{align}\label{eq:fermion_delta_r}
\begin{split}
  \deltauv_n(r)&:=\sum_{s=0}^r(-i\mstarsl\Prop{}{n})^{s}(-i\sigmafseuv(n)\Prop{}{n})(-i\mstarsl\Prop{}{n})^{r-s}\\[-0.4cm]&\qquad\qquad+\sum_{s=0}^{r-1}(-i\mstarsl\Prop{}{n})^{s}(-i\sigmasluv\Prop{}{n})(-i\mstarsl\Prop{}{n})^{r-1-s}\,,
\end{split}
\end{align}
while if $r=0$,
\begin{equation}
  \deltauv_n(0)=-i\sigmafseuv(n)\Prop{}{n}\,.
\end{equation}

In a similar way, the higher order versions of the background corrections to the fundamental emission one-loop result,  (\ref{eq:One_loop_01}), is for all $r\ge0$,
\begin{align}\label{eq:One_loop_rrp1} 
  \raisebox{-0.45cm}{\includegraphics{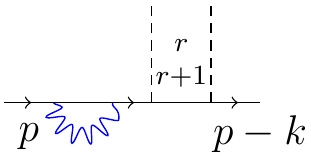}}&+\quad
  \raisebox{-0.45cm}{\includegraphics{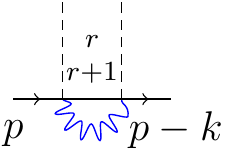}}+\quad
  \raisebox{-0.45cm}{\includegraphics{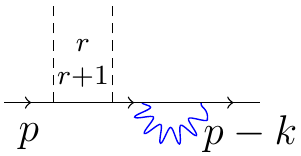}}\nonumber\\
  &\euv\Prop{}{-1}\deltauv_{-1}(r)\Out-\Out\Prop{}{0}\deltauv_0(r)\,.
\end{align} 
For the central terms, (\ref{eq:One_loop_11}), we get the generalisation, but now for $r\ge1$, that
\begin{align}\label{eq:One_loop_rr}
  \raisebox{-0.45cm}{\includegraphics{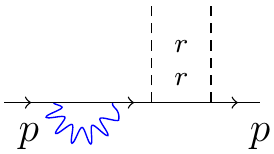}}&+\quad
  \raisebox{-0.45cm}{\includegraphics{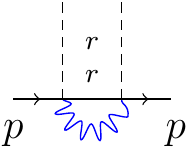}}+\quad
  \raisebox{-0.45cm}{\includegraphics{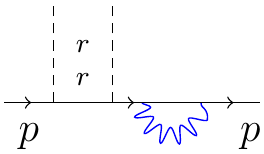}}\nonumber\\
  &\euv\In\Prop{}{-1}\deltauv_{-1}(r-1)\Out+\Prop{}{0}\deltauv_{0}(r)+\Out\Prop{}{1}\deltauv_{1}(r-1
  0)\In\,.
\end{align}
To this last result we also need to include  the $r=0$, single self-energy correction, given by $\Prop{}{0}\deltauv_0(0)$.

\bigskip

We now sum over all such sideband terms. In such a sum we need to remember that expression (\ref{eq:fermion_delta_r}) is only valid when $r\ge1$. The first sum here, though, does make sense when $r=0$, and gives the correct  result. The second sum in  (\ref{eq:fermion_delta_r}) only makes sense if $r\ge1$, but can be easily shifted to also start at $r=0$. Thus we get
\begin{align}
  \sum_{r=0}^{\infty}\Prop{}{n}\deltauv_n(r)&=\sum_{r=0}^{\infty}\sum_{s=0}^{r}\Prop{}{n}(-i\mstarsl\Prop{}{n})^{s}\big(-i(\sigmafseuv(n)+\sigmasluv)\big)\Prop{}{n}(-i\mstarsl\Prop{}{n})^{r-s}\nonumber\\
  &=\sum_{i=0}^{\infty}\Prop{}{n}(-i\mstarsl\Prop{}{n})^{i}\big(-i(\sigmafseuv(n)+\sigmasluv)\big)\sum_{j=0}^{\infty}\Prop{}{n}(-i\mstarsl\Prop{}{n})^{j}\nonumber\\
  &=\Prop{(m+\mstarsl)}{n}\big(-i(\sigmafseuv(n)+\sigmasluv)\big)\Prop{(m+\mstarsl)}{n}\,,
\end{align}
where in the last line we have used the matrix mass shifted, propagator result (\ref{eq:mass_shift2}).
 
\medskip
Applying this last result to the sums over the  sideband expressions (\ref{eq:One_loop_rp1r}), (\ref{eq:One_loop_rrp1}) and (\ref{eq:One_loop_rr}), gives us the fermionic, double line, one-loop result that
\begin{align} \label{eq:doubleline_loop2}
\raisebox{-0.45cm}{\includegraphics{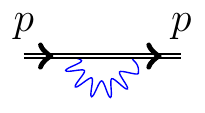}}&\euv 
\In\Prop{(m+\mstarsl)}{0}\big(-i(\sigmafseuv(0)+\sigmasluv)\big)\Prop{(m+\mstarsl)}{0}\nonumber\\
&\hspace{2cm}-\Prop{(m+\mstarsl)}{1}\big(-i(\sigmafseuv(1)+\sigmasluv)\big)\Prop{(m+\mstarsl)}{1}\In\nonumber\\ &\quad+\In\Prop{(m+\mstarsl)}{-1}\big(-i(\sigmafseuv(-1)+\sigmasluv)\big)\Prop{(m+\mstarsl)}{-1}\Out\nonumber\\
&\hspace{1.4cm}+\Prop{(m+\mstarsl)}{0}\big(-i(\sigmafseuv(0)+\sigmasluv)\big)\Prop{(m+\mstarsl)}{0}\\
&\hspace{2cm}+\Out\Prop{(m+\mstarsl)}{1}\big(-i(\sigmafseuv(1)+\sigmasluv)\big)\Prop{(m+\mstarsl)}{1}\In\nonumber\\ 
&\quad+\Prop{(m+\mstarsl)}{-1}\big(-i(\sigmafseuv(-1)+\sigmasluv)\big)\Prop{(m+\mstarsl)}{-1}\Out\nonumber\\
&\hspace{2cm}-\Out\Prop{(m+\mstarsl)}{0}\big(-i(\sigmafseuv(0)+\sigmasluv)\big)\Prop{(m+\mstarsl)}{0}. \nonumber
\end{align}
This last result, along with the tree level definition of the fermionic double line propagator, (\ref{eq:doubleline2}), gives a precise and surprisingly succinct description of the propagation in our background, to which standard renormalisation techniques can be applied.

\section{Renormalisation in a strong background}  \label{sec:renormalisation}
Having identified the one-loop ultraviolet poles for both scalar and fermionic matter in our background, we can now introduce counterterms and discuss the renormalisation of these theories. Surprisingly, we  shall see that there are very simple interpretations of the renormalisation process in both cases.

\medskip 

The one-loop corrections to the double line  propagators, introduced here in sections~\ref{sec:scalar_matter} and~\ref{sec:strong_fermion_loop}, now need to be interpreted in terms of bare fields and then renormalised via the introduction of appropriate counterterms. In \cite{Lavelle:2019lys} and \cite{Lavelle:2019vcz}, it was seen that, even in a weak background, an additional counterterm was needed to renormalise the fermionic theory.   Having extended here these  one-loop calculations, for both types of matter, to all orders in the background, we can now look more precisely at the set of counterterms needed to renormalise the two theories.

\medskip

Starting with scalar matter,  the expressions for the self-energy, (\ref{eq:sigma_scalar}), and background induced mass correction, (\ref{eq:sigma_v}), need to be added together and properly interpreted before we introduce  any counterterms. We can write these  one-loop corrections as 
\begin{equation}\label{eq:scalar_se_mstar1}
  -i\big(\sigmaseuv+\sigmamauv\big)=-i\frac{e^2}{(4\pi)^2}\frac1\varepsilon\Big(3m^2-i(\xi-3)\Props{m^2}^{-1}+\xi\smstar^2\Big)\,.
\end{equation}
The first thing to note about this expression is that the inverse propagator term here does not match the double line propagators that surround these corrections in (\ref{eq:scalar_doubleline_loop}). Under the simple mass shift,  $\Props{m^2}^{-1}=\Props{m^2+\smstar^2}^{-1}-i\smstar^2$, we get the full inverse propagator plus additional $\smstar^2$ terms that now combine, in a very attractive way, with the mass term to give
\begin{equation}\label{eq:scalar_se_mstar2}
  -i\big(\sigmaseuv+\sigmamauv\big)=-i\frac{e^2}{(4\pi)^2}\frac1\varepsilon\Big(3(m^2+\smstar^2)-i(\xi-3)\Props{m^2+\smstar^2}^{-1}\Big)\,.
\end{equation}
Note that this last expression can be interpreted as the usual ultraviolet pole of the self-energy for a scalar particle of mass $m_*^2=m^2+\smstar^2$.  Now introducing the tree-level, but not free,  bare mass $m^2_*$ and the, Volkov field,  wave function renormalisation, allows for the expansion in terms of mass, $\delta_{m^2_*}$, and Volkov wave function, $\delta_{2}$, counterterms so that
\begin{equation}
  -i\big(\sigmaseuv+\sigmamauv\big)\to-i\big(\sigmaseuv+\sigmamauv+m^2_*\delta_{m^2_*}-i\Props{m^2_*}^{-1}\delta_2\big)\,.
\end{equation}
From (\ref{eq:scalar_se_mstar2}) we can now read off the strong field renormalisation conditions that
\begin{equation}\label{eq:scalar_renormalisation}
  \delta_{m^2_*}=-\frac{e^2}{(4\pi)^2}\frac3{\varepsilon}\qquad\mathrm{and}\qquad
  \delta_2=-(\xi-3)\frac{e^2}{(4\pi)^2}\frac1{\varepsilon}\,.
\end{equation}
The most striking thing about this result is its simplicity. In terms of the strong field variables,  $m^2_*$ and $\Props{m^2_*}^{-1}$, we have the same multiplicative structure familiar from the scalar matter in a vacuum. If we were not in the momentum gauge, then many additional gauge artifacts would obstruct this simple result.  

\medskip

For the fermionic theory, this argument needs to be repeated in each of the sidebands, with propagator $\Prop{}{n}$, for $n$ either $0$ or $\pm1$. We quickly see that for the $n^\mathrm{th}$ such sideband,
\begin{equation}\label{eq:fermionic_se_mstar1}
  -i\big(\sigmafseuv(n)+\sigmasluv\big)=-i\frac{e^2}{(4\pi)^2}\frac1\varepsilon\Big(3m-i\xi\Prop{(m)}{n}^{-1}+\xi\mstarsl\Big)\,.
\end{equation}
Again, the inverse propagator here needs to match the terms multiplying it, so we make the replacement $\Prop{(m)}{n}^{-1}=\Prop{(m+\mstarsl)}{}^{-1}-i\mstarsl$. The end result of this shift is that 
\begin{equation}\label{eq:fermionic_se_mstar2}
  -i\big(\sigmafseuv(n)+\sigmasluv\big)=-i\frac{e^2}{(4\pi)^2}\frac1\varepsilon\Big(3m-i\xi\Prop{(m+\mstarsl)}{n}^{-1}\Big)\,.
\end{equation} 
In contrast to the scalar result above, now we see only a renormalisation of the vacuum mass, $m$, but the wave function term is still with respect to the full, strong field normalisation.

Thus, in this fermionic theory, we need to introduce just a vacuum mass counterterm, $\delta_{m}$, along with the strong field,  wave function term, $\delta_2$. Then, written in terms of renormalised fields, we have   
\begin{equation}
  -i\big(\sigmafseuv(n)+\sigmasluv\big)\to-i\big(\sigmafseuv(n)+\sigmasluv+m\delta_{m}-i\Prop{(m+\mstarsl)}{n}^{-1}\delta_2\big)\,.
\end{equation}
The  renormalisation conditions that follow from this and  (\ref{eq:scalar_se_mstar2}) are now 
\begin{equation}\label{eq:fermion_renormalisation}
  \delta_{m}=-\frac{e^2}{(4\pi)^2}\frac3{\varepsilon}\qquad\mathrm{and}\qquad
  \delta_2=-\xi\frac{e^2}{(4\pi)^2}\frac1{\varepsilon}\,.
\end{equation}

\medskip

The differences revealed in this section in the  renormalisation conditions needed for scalar, (\ref{eq:scalar_renormalisation}), and fermionic,  (\ref{eq:fermion_renormalisation}), matter are surprising. That the gauge fixing conditions enter differently is not itself unexpected, but  the fact that different classes of counterterms are needed seems unexpected. In particular, the contrast between the full mass counter term for scalars and only the vacuum mass counter term for fermions  seems unexpected. It is not clear, a priori, why this should be the case. Especially given the fact that both types of matter have the same, strong field,  counterterm structure for the wave function renormalisation associated with the full propagators (\ref{eq:scalar_mass_shift}) and (\ref{eq:doubleline2}).

\section{Conclusions}  
There is huge theoretical and experimental interest in particle physics in an intense laser background. Much of the theoretical work builds upon the Volkov solution, where there is known to be a mass shift and a loss of translational invariance, both induced by the background. The Volkov solution has built into it a gauge freedom. In this paper we have introduced an additional gauge fixing condition on the background, which we call the momentum gauge, which dramatically simplifies the description of charged matter   propagation. 

For scalar matter, in a circularly polarised background, we have seen in this paper that only one type of background interaction survives in this gauge. This interaction respects translation invariance, and can be easily summed to all orders. As the background non-translational invariance  has been gauged away, all the familiar tools from vacuum scalar QED could be deployed. The strong field solution developed here exhibits the background induced  mass shift, but none of the normally expected sideband structures, which are revealed here to be gauge artefacts, even at one-loop.

For fermionic matter, a small number of  sidebands persist in the momentum gauge. This corresponds to a limited violation of translational invariance. Despite this, we have been able to develop momentum space techniques to construct the propagator and its one-loop, ultraviolet corrections. The background induced mass term has a matrix structure that is common to each sideband. We emphasise that in this momentum gauge, the infinite tower of sidebands has reduced to just seven terms for fermionic matter and only one for the scalar theory. 

Our analysis of the renormalisation has further revealed a difference in the counterterm structures needed in both theories. For the scalar matter, (\ref{eq:scalar_se_mstar2}), we have renormalisation of the shifted mass for the full propagator, and of the residue of the pole at this shifted mass. For fermionic matter, (\ref{eq:fermionic_se_mstar2}),  the vacuum mass is renormalised, rather than the shifted mass,  but we saw that it is the residue of the shifted mass that acquires a wave function  renormalisation.

Our focus on circular polarisation for the background field has led to particularly simple results for both the tree level and one-loop renormalisation of these theories. The most immediate impact of widening the class of polarisations is that the terms $v$ and $v^*$, as defined in (\ref{eq:v_vanish}), no longer vanish, and one will get Bessel functions of these terms as factors in the sideband structure. This was alluded to in  equation  (\ref{eq:simp1}) of this paper. These added effects from the background  will impact on our results for  both scalar and fermionic matter. However, the evidence from \cite{Lavelle:2019lys} and \cite{Lavelle:2019vcz} is that these terms do not acquire  any one-loop corrections. We conjecture that this observation will also hold in a strong background for both types of matter. The circular polarisation case considered in this paper seems to represent the simplest configuration that captures the essential physics of the loop corrections in a plane wave background, for both types of matter. 

Future work will also include an analysis of the finite parts and the infrared structures associated with  charged matter  propagating in the plane wave background. For scattering processes the momentum gauge can be applied to one leg, thus simplifying some of the one-loop structures, and the details of this will be presented elsewhere.

\section{Acknowledgements} 
We thank Tom Heinzl for discussions.

\newpage

\appendix

\section{Perturbative factorisation results}\label{sec:appendix1}
In this appendix we collect together the details of the arguments that lead to the key perturbative factorisation result (\ref{eq:factorise}), and then the  explicit expressions  (\ref{eq:ind1}), (\ref{eq:ind2}) and (\ref{eq:ind3}) that follow from it. The simplicity of these  results all depend critically on our choice of gauge and polarisation.  

Although the factorisation result has been stated quite generally in (\ref{eq:factorise}), the vanishing result (\ref{eq:n1_in_n2_out_zero}) means that we only need to consider three non-trivial cases corresponding to a net absorption ($r_2=r_1-1$), a balanced interaction ($r_2=r_1$) or a net emission ($r_2=r_1+1$). 

A net absorption means that there will be one extra power of the absorption vertex (\ref{eq:Ab}) over the emission vertex (\ref{eq:Em}). The vanishing results (\ref{eq:simp_AandEwithP}) means that these vertices must alternate and hence, for $r\ge0$,
 \begin{equation}\label{eq:rinrminus1out}
  \raisebox{-0.54cm}{\includegraphics{fermion_r_plus1_in_r_out.pdf}} =\Prop{}{1}\Ab(\Prop{}{0}\Em\Prop{}{1}\Ab)^r\Prop{}{0}\,.
\end{equation} 
If $r=0$ this reduces to the fundamental absorption process (\ref{eq:1_in_0_out}), while if $r>0$ we have, using the vanishing identities (\ref{eq:simp_AandEwithP}),
\begin{align}
 \Prop{}{1}\Ab(\Prop{}{0}\Em\Prop{}{1}\Ab)^r\Prop{}{0}&=\Prop{}{1}\Ab(\Prop{}{0}\Em\Prop{}{1}\Ab)^{r-1}(\Prop{}{0}\Em\Prop{}{1}\Ab+\Prop{}{0}\Ab\Prop{}{-1}\Em)\Prop{}{0}\nonumber \\
 &=\Prop{}{1}\Ab(\Prop{}{0}\Em\Prop{}{1}\Ab)^{r-1}\Prop{}{0}(\Em\Prop{}{1}\Ab+\Ab\Prop{}{-1}\Em)\Prop{}{0}\,.
\end{align} 
Hence we recover the factorisation identity that, for $r>0$,
\begin{equation}\label{eq:fac_ab}
 \raisebox{-0.54cm}{\includegraphics{fermion_r_plus1_in_r_out.pdf}} =
  \raisebox{-0.54cm}{\includegraphics{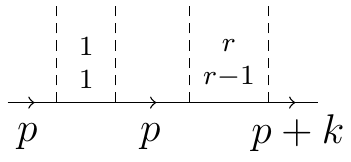}} \,.
\end{equation} 
The dual emission version of this factorisation identity can be shown in a very similar way. For the balanced case there are now two contributing terms when we have $r>0$ absorptions and emissions: $(\Prop{}{0}\Em\Prop{}{1}\Ab)^r\Prop{}{0}+(\Prop{}{0}\Ab\Prop{}{-1}\Em)^r\Prop{}{0}$. But, using the vanishing identities again, this can be written as $(\Prop{}{0}\Em\Prop{}{1}\Ab+\Prop{}{0}\Ab\Prop{}{-1}\Em)^r\Prop{}{0}$, from which the factorisation result immediately follows.
 
These factorisation results now allow for an inductive derivation of the key identities (\ref{eq:ind1}), (\ref{eq:ind2}) and (\ref{eq:ind3}), where the base cases have already been seen in  (\ref{eq:1_in_0_out}), (\ref{eq:0_in_1_out}) and (\ref{eq:1_in_1_out}). In fact, we only need to show (\ref{eq:ind3}), as the other two then then follow using repeated applications of the factorisation results.

Assuming the identity (\ref{eq:ind3}) holds for $r\ge1$ absorptions and emissions, the factorisation result  then allows us to write
\begin{align}\label{eq:induction_id3}
 \raisebox{-0.5cm}{\includegraphics{fermion_r_plus1_in_r_plus1_out.pdf}} 
 &=\raisebox{-0.45cm}{\includegraphics{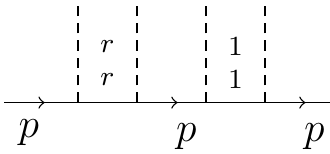}} \nonumber\\
  &=\big(\In\Prop{}{-1}\Out+\Prop{}{0}(-i\mstarsl\Prop{}{0})+\Out\Prop{}{1}\In\big)\Prop{}{0}^{-1}\\
  &\qquad\times\big(\In\Prop{}{-1}(-i\mstarsl\Prop{}{-1})^{r-1}\Out+\Prop{}{0}(-i\mstarsl\Prop{}{0})^r+\Out\Prop{}{1}(-i\mstarsl\Prop{}{1})^{r-1}\In\big)\,. \nonumber
\end{align}  
Expanding the right hand side of (\ref{eq:induction_id3}) yields nine potential terms but this  quickly reduce to three by using the trivial identities that $\mstarsl\In=\mstarsl\Out=0$, along with (\ref{eq:simp_IProp}). Thus we get
\begin{align}\label{eq:induction_id2}
\begin{split}
  \raisebox{-0.5cm}{\includegraphics{fermion_r_plus1_in_r_plus1_out.pdf}} 
 &=\In\Prop{}{-1}\Out\Prop{}{0}^{-1}\In\Prop{}{-1}(-i\mstarsl\Prop{}{-1})^{r-1}\Out+\Prop{}{0}(-i\mstarsl\Prop{}{0})^{r+1}\\&\qquad+\Out\Prop{}{1}\In\Prop{}{0}^{1}\Out\Prop{}{1}(-i\mstarsl\Prop{}{1})^{r-1}\In\,.
\end{split} 
\end{align}
Now we can use the identities  (\ref{eq:in_iprop_out}) and (\ref{eq:in_Propntor_in})  to deduce the claimed result (\ref{eq:ind3}), for all $r\ge0$. 

\section{Derivation of Ritus matrices} \label{sec:app_ritus}

The compact tree level result (\ref{eq:doubleline2}) is of interest in its own right since there are other approaches to this double line propagator that are not based on perturbation theory, and look very different. As a check of the results developed here, we now  show how the more familiar Ritus matrices \cite{Ritus:1972ky} reduce to the terms in  (\ref{eq:doubleline2}) for our choice of circular polarisation and use of the momentum gauge.

In order to trace the consequences of the assumptions made in this paper, we first consider the more general elliptic class of polarisations, which includes circular and linear polarisation as limiting cases. In equation (44) of \cite{Lavelle:2017dzx}, a suitable time-ordered product of Volkov fields was calculated and, written here in terms of a double line, shown to be equal to 
\begin{equation}\label{eq:doubleline3}
  \raisebox{-0.05cm}{\includegraphics{fermion_doubleline.pdf}}=\sum_{n,r}\ee^{i(n+r)\xk} J_{n+r}(\Omega'_1,v',\Omega'_2)\Prop{(m+\mstarsl)}{n}\overline{J}_n(\Omega'_1,v',\Omega'_2)\ee^{-in\xk}\,.
\end{equation} 
The notation used here is more refined than that used in \cite{Lavelle:2017dzx}, and is essentially found in the discussion leading to  equation (66) of \cite{Lavelle:2019lys}. The normalising functions,  $J_n$, are the elliptic class of generalised Bessel functions with first and last arguments \hbox{$\Omega'_1=-(\In'+\Out')$} and $\Omega'_2=-i(\In'-\Out')$. The \lq In\rq\ and \lq Out\rq\ terms here are written with primes to signify that they do not include the final exponential factors in the complex potential~(\ref{eq:A_circular}), used in this paper. But we note that the terms in (\ref{eq:doubleline3}) are multiplied by the exponential $\ee^{ir\xk}$, which we have factorised to match the order of the normalising Bessel functions. These factors  can then be reabsorbed into the arguments of the Bessel functions, with the result that the parameters shift from functions of $\In'$ and $\Out'$ to those of $\In$ and $\Out$, as used in this paper. In the same way, the scalar $v'$ becomes the $v$ of equation (\ref{eq:v_vanish}),  which only vanishes for circular polarisation. By shifting the background induced mass term out of the propagator in (\ref{eq:doubleline3}), one recovers the Ritus representation of the double line propagator.  See the discussion around equation (35) in \cite{Lavelle:2017dzx} for more details.
 
We now start to restrict this quite general representation for the double line propagator to the situation considered in this paper. The first thing to note is that  in the momentum gauge  the generalised Bessel functions can be easily expanded in terms of the \lq In\rq\ and \lq Out\ matrices resulting in the simplification that  
\begin{equation}\label{eq:simp1}
   J_n\big(-(\In+\Out),v,-i(\In-\Out)\big)=\begin{cases}
    J_{(n+1)/2}(v)\In-J_{(n-1)/2}(v)\Out, &\text{ if $n$ is odd }\\[0.25cm]
    J_{n/2}(v), &\text{ if $n$ is even, }
    \end{cases}
\end{equation}
where the right hand side now involves just standard Bessel functions.

If we now impose the circular polarisation condition that $v=0$, then these Bessel functions further simplify so that only three terms survive:
\begin{equation}
  J_{-1}\big(-(\In+\Out),0,-i(\In-\Out)\big)=\In\,,
\end{equation}
\begin{equation}
  J_{0}\big(-(\In+\Out),0,-i(\In-\Out)\big)=1
\end{equation}
and
\begin{equation}
  J_{1}\big(-(\In+\Out),0,-i(\In-\Out)\big)=-\Out\,.
\end{equation}
Inserting these results into the expression (\ref{eq:doubleline3}) then quickly recovers  the result (\ref{eq:doubleline2}) for the double line propagator that was derived perturbatively in this paper.

\section{Induced mass loop correction} \label{sec:app_mass}
In this appendix we wish to sketch the key steps in deriving the one-loop result~(\ref{eq:1_1_loop}) and hence the background induced mass loop correction (\ref{eq:massUV}).

We have seen that the background induced mass arises  from the mixture of absorptions and emissions to the background. The expression underlying this sideband result (\ref{eq:1_in_1_out}) is
\begin{equation}\label{eq:one_in_one_out} 
\raisebox{-0.5cm}{\includegraphics{fermion_one_in_one_out.pdf}}=\Prop{}{0}\big(\Ab\Prop{}{-1}\Em+\Em\Prop{}{1}\Ab\big)\Prop{}{0}\,.
\end{equation} 
Following the discussion related to figure 9 in \cite{Lavelle:2019lys}, the ultraviolet pole of the one-loop version of the bracketed expression on the right in (\ref{eq:one_in_one_out}) can easily be found by the simple algebraic replacement:  $\Prop{}{n}\to\Prop{}{n}+\Prop{}{n}(-i\sigmafseuv(n))\Prop{}{n}$; $\Ab\to\Ab-i\sigmauv_{\mathrm{in}}$ and $\Em\to\Em-i\sigmauv_{\mathrm{out}}$. Hence we have the one-loop ultraviolet pole identification that
\begin{align}
\begin{split}
  \raisebox{-0.5cm}{\includegraphics{fermion_laser_line_loop_11.pdf}}&\euv\Prop{}{0}\big(-i\sigmauv_{\mathrm{in}}\Prop{}{-1}\Em+\Ab\Prop{}{-1}(-i\sigmafseuv(-1))\Prop{}{-1}\Em+\Ab\Prop{}{-1}(-i\sigmauv_{\mathrm{out}})\\
  &\hspace{1.4cm}-i\sigmauv_{\mathrm{out}}\Prop{}{1}\Ab+\Em\Prop{}{1}(-i\sigmafseuv(1)\Prop{}{1}\Ab+\Em\Prop{}{1}(-i\sigmauv_{\mathrm{in}})\big)\Prop{}{0}\,.
  \end{split}
\end{align}
These can now be evaluated in terms of sideband loop corrections using the tree-level identities in (\ref{eq:SbP})  and the  vertex results (\ref{eq:fermion_ab_as_se_uv}) and (\ref{eq:out_loop}).

One quickly finds that 
\begin{align}\label{eq:nearly_1_1}
\begin{split}
  \raisebox{-0.45cm}{\includegraphics{fermion_laser_line_loop_11.pdf}}&\euv\In\Prop{}{-1}(-i\sigmafseuv(-1))\Prop{}{-1}\Out\\[-2mm]&\hspace{0.5cm}-\Prop{}{0}\Big(\In(-i\sigmafseuv(-1))\Out+\Out(-i\sigmafseuv(1))\In\Big)\Prop{}{0}\\&\hspace{1cm}+\Out\Prop{}{1}(-i\sigmafseuv(1))\Prop{}{1}\In\\&\hspace{1.5cm}-\Prop{}{0}(-i\sigmafseuv(0))\big(\In\Prop{}{-1}\Out+\Out\Prop{}{1}\In\big)\\&\hspace{1.5cm}-\big(\In\Prop{}{-1}\Out+\Out\Prop{}{1}\In\big)(-i\sigmafseuv(0))\Prop{}{0}\,.
  \end{split}
\end{align}
Now using the results (\ref{eq:in_iprop_out}) and (\ref{eq:sigma_uvse_fermion}), we see that 
\begin{equation}
  \In(-i\sigmafseuv(-1))\Out+\Out(-i\sigmafseuv(1))\In=\frac{ie^2}{(4\pi)^2}\xi\mstarsl\frac1\varepsilon\,.
\end{equation}
Combining this last expression with the expansion (\ref{eq:nearly_1_1}) then gives the key result (\ref{eq:1_1_loop}) in the main text.

\section{Some strong field results} \label{sec:app_strong}
The tree level, fermionic  factorisation identity (\ref{eq:factorise}) immediately gives the ultraviolet, one-loop, factorisation result that
\begin{equation}\label{eq:loop_factorise} 
\raisebox{-0.4cm}{\includegraphics{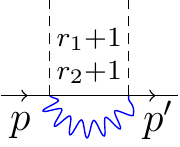}}\euv
\raisebox{-0.5cm}{\includegraphics{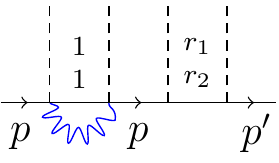}}+
\raisebox{-0.5cm}{\includegraphics{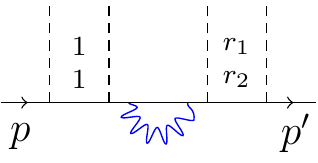}}+
\raisebox{-0.5cm}{\includegraphics{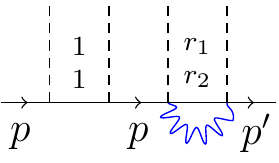}}
\,.
\end{equation}   
This now allows us to inductively build up  the one-loop corrections to all the higher order interactions with the background. 

For the central terms, where $r_1=r_2$, we now use this ultraviolet factorisation  result to prove that, for all $r\ge1$,
\begin{align}\label{eq:factorisation1}
  &\raisebox{-0.45cm}{\includegraphics{fermion_rr_in_loop.pdf}}\euv\In\Prop{}{-1}\deltauv_{-1}(r-1)\Out+\Prop{}{0}\deltauv_{0}(r)+\Out\Prop{}{1}\deltauv_{1}(r-1)\In \\
  &\quad-\big(\In\Prop{}{-1}(-i\mstarsl\Prop{}{-1})^{r-1}\Out+\Prop{}{0}(-i\mstarsl\Prop{}{0})^{r}+\Out\Prop{}{1}(-i\mstarsl\Prop{}{1})^{r-1}\In\big)(-i\sigmafseuv(0))\Prop{}{0}\nonumber\\
  &\quad-\Prop{}{0}(-i\sigmafseuv(0))\big(\In\Prop{}{-1}(-i\mstarsl\Prop{}{-1})^{r}\Out+\Prop{}{0}(-i\mstarsl\Prop{}{0})^{r+1}+\Out\Prop{}{1}(-i\mstarsl\Prop{}{1})^{r}\In\big)\,,\nonumber
\end{align}
which is equivalent to (\ref{eq:One_loop_rr}).

When $r=1$, this expression reduces  to our earlier one-loop calculation, (\ref{eq:1_1_loop}). To then show the result when we have  $r+1$ absorptions and emissions, we use the factorisation identity to disentangle the interactions, thus reducing to three processes where we have less interactions spanned by the loop, and hence the inductive assumption can be used. 

So the key  factorisation result we need is that
\begin{equation}\label{eq:factorisation_central}
  \raisebox{-0.45cm}{\includegraphics{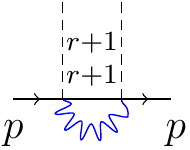}}\euv\raisebox{-0.5cm}{\includegraphics{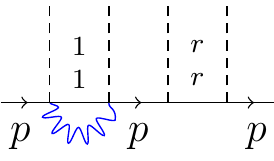}}+\raisebox{-0.5cm}{\includegraphics{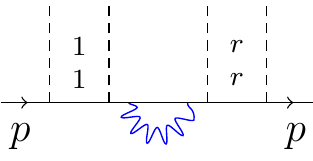}}+
\raisebox{-0.5cm}{\includegraphics{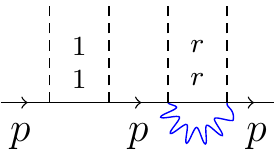}}\,.
\end{equation}
The second two diagrams here can be combined and simplified, after repeated use of the identities (\ref{eq:in_Propntor_in}), to give
 \begin{align}\label{eq:factorisation2}
  &\raisebox{-0.5cm}{\includegraphics{fermion_loop_factorisation4.pdf}}+
\raisebox{-0.5cm}{\includegraphics{fermion_loop_factorisation5.pdf}}\\&\euv\In\Prop{}{-1}\deltauv_{-1}(r-1)(-i\mstarsl\Prop{}{-1})\Out+\Prop{}{0}\deltauv_{0}(r)(-i\mstarsl\Prop{}{0})+\Out\Prop{}{1}\deltauv_{1}(r-1)(-i\mstarsl\Prop{}{1})\In \nonumber\\
  &\quad-\Prop{}{0}(-i\sigmafseuv(0))\big(\In\Prop{}{-1}(-i\mstarsl\Prop{}{-1})^{r}\Out+\Prop{}{0}(-i\mstarsl\Prop{}{0})^{r+1}+\Out\Prop{}{1}(-i\mstarsl\Prop{}{1})^{r}\In\big)\,,\nonumber
\end{align}
From the definition (\ref{eq:fermion_delta_r}), we quickly see that 
\begin{align}
\begin{split}
  \deltauv_{n}(r)(-i\mstarsl\Prop{}{n})= \deltauv_{n}(r+1)&-(-i\mstarsl\Prop{}{n})^{r+1}(-i\sigmafseuv(n)\Prop{}{n})\\
  &-(-i\mstarsl\Prop{}{n})^{r}(-i\sigmasluv\Prop{}{n})\,.
\end{split}
\end{align}
Using this result in (\ref{eq:factorisation2}), we find that
 \begin{align}\label{eq:factorisation3}
  &\raisebox{-0.5cm}{\includegraphics{fermion_loop_factorisation4.pdf}}+
\raisebox{-0.5cm}{\includegraphics{fermion_loop_factorisation5.pdf}}\\&\euv\In\Prop{}{-1}\deltauv_{-1}(r)\Out+\Prop{}{0}\deltauv_{0}(r+1)+\Out\Prop{}{1}\deltauv_{1}(r)\In -\raisebox{-0.45cm}{\includegraphics{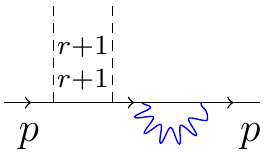}}\nonumber\\
  &\qquad+\mathrm{terms\ linear\ in\ } \deltauv_{n}(0)\ \mathrm{and\ }\deltauv_{0}(1)\,.\nonumber 
\end{align}
When the loop straddles the initial absorption and emission process, we get
\begin{align}\label{eq:factorisation4}
  &\raisebox{-0.5cm}{\includegraphics{fermion_loop_factorisation6.pdf}}\euv -\raisebox{-0.45cm}{\includegraphics{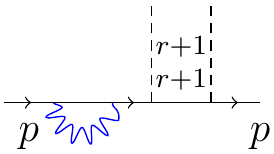}}-\mathrm{terms\ linear\ in\ } \deltauv_{n}(0)\ \mathrm{and\ }\deltauv_{0}(1)\,,
\end{align} 
where the final terms here are, up to a sign, the same as in (\ref{eq:factorisation3}). 

Combining the expressions (\ref{eq:factorisation3}) and (\ref{eq:factorisation4}), we see that 
 \begin{align}\label{eq:factorisation5}
 \begin{split}
   &\raisebox{-0.5cm}{\includegraphics{fermion_rp1rp1_in_loop.pdf}}\euv\In\Prop{}{-1}\deltauv_{-1}(r)\Out+\Prop{}{0}\deltauv_{0}(r+1)+\Out\Prop{}{1}\deltauv_{1}(r)\In \\&\hspace{3cm}-\raisebox{-0.45cm}{\includegraphics{fermion_rp1rp1_then_loop.pdf}}-\raisebox{-0.45cm}{\includegraphics{fermion_loop_then_rp1rp1.pdf}}\,.
 \end{split}
  \end{align} 
Thus the one-loop, central sideband result (\ref{eq:factorisation1}) holds for all $r\ge1$. 
 
\bigskip

Armed with this central result, we can rapidly derive the expressions for the upper and lower sidebands,(\ref{eq:One_loop_rp1r}) and (\ref{eq:One_loop_rrp1}), by using, for example, the ultraviolet, one-loop factorisation 
\begin{equation}
  \raisebox{-0.45cm}{\includegraphics{fermion_rp1r_in_loop.pdf}}
  \euv
\raisebox{-0.5cm}{\includegraphics{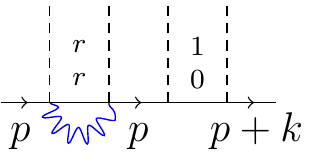}}+
\raisebox{-0.5cm}{\includegraphics{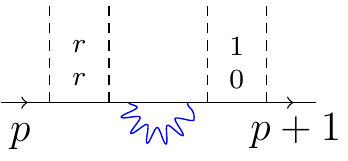}}+
\raisebox{-0.5cm}{\includegraphics{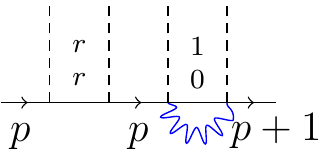}}
\,.
\end{equation}


\end{document}